\documentclass{aa}  

\usepackage{graphicx}
\usepackage{grffile}
\usepackage{natbib}
\usepackage{txfonts}
\usepackage{lastpage}
\usepackage{hyperref}
\usepackage{comment}

\newcommand{\orcid}{\href{https://orcid.org/0000-0001-8824-9385}
{\protect\includegraphics[width=8pt]{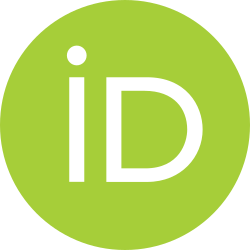}}}

\begin{document} 
\title{Low-Frequency Stochastic Gravitational-Wave Background in \\ \textit{Gaia} DR3 catalog}
\titlerunning{Stochastic Gravitational-Wave Background in \textit{Gaia} data}
    \author{ V. Akhmetov\inst{1,2,3} \orcid , L. Filipello\inst{1,4}, M. Crosta\inst{1}, M. G. Lattanzi\inst{1}, B. Bucciarelli\inst{1}, U. Abbas\inst{1} \and F. Santucci\inst{1}
        }
    \authorrunning{Akhmetov et al.}

   \institute{
   $^{1}$ INAF - Astronomical Observatory of Torino, via Osservatorio 20, I-10025, Turin, Italy
   \\ \email{\href{mailto:akhmetovvs@gmail.com}{akhmetovvs@gmail.com}}\\   
   $^{2}$ Main Astronomical Observatory, National  Academy of Sciences of Ukraine,
           27 Akademika Zabolotnoho St, 03143 Kyiv, Ukraine\\       
    $^{3}$ Institute of Astronomy of V.N.Karazin Kharkiv National University, Svobody sq. 4, 61022, Kharkiv, Ukraine \\
    $^{4}$ University of Turin, Department of Physics, Via Pietro Giuria 1, Turin, Italy\\
   }

 
  \abstract
  {In the era of Gravitational Waves observations, searching signals at the very low-frequencies ($<10^{-7}$ Hz) can bring important constraints on cosmology and the early universe. As the next \textit{Gaia} Data Release (DR4) approaches, high-precision astrometry is emerging as a powerful complement to other methods, such as pulsar timing arrays (PTAs), in the search for gravitational wave backgrounds and discrete sources.}
  {We investigate the potential to detect low-frequency gravitational waves (GWs) through their imprints on the proper motions of distant quasars observed by the \textit{Gaia} mission. Using astrometric data from \textit{Gaia} DR3, we simulate the effect of GWs on the proper motions of quasars, incorporating their actual sky positions and measurement uncertainties.}
  {We investigate two closely related data analysis techniques used for the extraction and characterization of GW signals from quasar proper motions: Vector Spherical Harmonics (VSH) and angular correlation functions, commonly referred to as Hellings–Downs curves (HDC). Using realistic simulated data, we forecast their sensitivity and accuracy to GWs, and evaluate the impact systematic errors might have on them. From these simulations, we derive an upper limit on the amplitude of a stochastic GW background, constrained by the observational timespan, astrometric precision, and the sky distribution of the quasars. Also, we test scalability, as a significantly growing number of quasars with high quality proper motions is expected in the next \textit{Gaia} data release (DR4).}
  {VSH decomposition appears less sensitive to uneven sky sampling and anisotropic noise. The HDC approach retains a larger fraction of the pairwise correlation information and therefore exhibits higher raw statistical sensitivity under idealized conditions. However, this increased sensitivity comes at the cost of stronger susceptibility to anisotropic sky coverage, correlated noise, and substantially higher computational complexity scales as $N^2$.} 
  {We find that the sensitivity floor for a detectable GW strain is $10^{-11}$ for \textit{Gaia} DR3 proper motion errors, with possible improvements to $\sim 3\times10^{-12}$ for the next \textit{Gaia} Data Release 4 (for the same number of quasars). This limit holds for a stochastic spectrum of GW integrated over all frequencies less than half the inverse of the 34 month span of the observations \textit{Gaia} DR3, or about $f_{GW} \lesssim $ 5.6 nHz. We investigated how different data-restriction and weighting schemes influence the final estimate of the GW strain.}

   \keywords{  gravitational waves - astrometry - proper motion - methods: numerical - methods: observational - Gaia }

   \maketitle
%
\section{Introduction}
After entering the era of Gravitational Wave (GW) astronomy, it has become crucial to leverage expertise from different fields to better characterize these weak-regime gravitational phenomena and to develop complementary detection techniques. Although GWs were not the primary target of the \textit{Gaia} mission \citep{Gaia:2021gsq}, its high-precision astrometry provides a valuable resource for detecting signals in the very-low-frequency regime.

A key aspect of GW detection through astrometry is that, while Pulsar Timing Arrays (PTAs) -- the primary experimental method for low-frequency GW detection -- are sensitive only to the changes in pulse arrival times caused by GWs along the line of sight (effectively a radial displacement of the pulsar position as projected along the observer’s line of sight), astrometric observations can probe the transverse component of GW perturbations, i.e., the angular motion on the celestial sphere \citep{Hobbs_2010, Gwinn_1997}. Moreover, because proper motions are secular measurements, the accessible frequency range is bounded by the duration of the mission (upper limit) and the look-back time of the observed quasars (lower limit), which is strongly dependent on cosmology \citep{Darling_2025}. This will be further discussed in the following sections.

Assuming that GW induced angular displacements are the sole contributor to quasar proper motions, it is necessary to employ methods capable of extracting these subtle signals. Two main approaches have been proposed: (i) analysis of angular motions using Vector Spherical Harmonics (VSH) to probe different multipoles \cite{Vityazev_2009, Mignard:2012xm}, and (ii) computation of correlations following a Hellings–Downs (HD) for GW induced perturbations, examining all possible quasar pairs at varying angular separations \cite{HellingsDowns1983, Gwinn_1997}.

High-precision astrometry complements existing GW detection techniques by providing an independent probe of low-frequency signals. GWs induce deflections in the positions of distant quasars, producing correlated apparent proper motions across the sky. The amplitude of the GW strain detectable by astrometry is primarily limited by the astrometric precision, the duration of the observations, and the sky coverage of the quasar sample \citep{Jaraba2023, Darling_2025}.

In this work, our aim is to develop a fast and highly parallelized tool for the simulation and analysis of GW induced astrometric deflections. With future $Gaia$ data releases, the number of quasars with proper motions minimally affected by systematics is expected to increase significantly. Consequently, the computational cost will increase for both analysis techniques: Hellings--Downs Correlation (HDC) analysis scales approximately as $\mathcal{O}(N^2)$, where $N$ is the number of quasars, whereas higher-order VSH multipoles scale linearly with $N$.
Furthermore, VSH decomposition is found to be statistically robust, less sensitive to uneven quasar sampling, and computationally more efficient than HDC methods. 

In addition, in this weak-regime it is challenging to distinguish systematic errors from genuine physical effects. Therefore, quasars can also serve as a tool to characterize and quantify the magnitude of the $Gaia$ systematics \citep{Lindegren2021}.

In Sect. \ref{sec:methods}, we first summarize the theoretical tools used to extract and characterize stochastic GW signals from proper motions, followed by a description of the computational techniques adopted for noise reduction and data compression.

In Sect. \ref{sec:simulation}, we evaluate the effects on proper motions due to different types of GW sources on simulated data and examine the possible effects of systematics and anisotropies on the analysis to forecast the possible intrinsic limits of astrometric data.

In Sect. \ref{sec:results}, we present the data analysis performed on real quasars from $Gaia$-CRF3 \citep{Klioner2022} to estimate current systematic errors and to obtain updated constraints on the low-frequency Gravitational Wave Background.

\section{Methods}
\label{sec:methods}

Astrometric deflections induced by GWs can be characterized as angular displacements of distant sources on the celestial sphere. In the far-away approximation, for an observer at rest (which in the real scenario will be the barycenter of the Solar System), the astrometric displacement of the unitary direction $u^i$ of a source in the celestial sphere due to a passing plane GW can be expressed as
\begin{equation}
    \delta u^i = \frac{u^i + p^i}{2 (1 + \mathbf{u} \cdot \mathbf{p})}h_{jk} u^j u^k - \frac{1}{2} h_{ij} u^j,
\end{equation}
where vector $\mathbf{p}$ is the propagation direction of the wave, vector $\mathbf{u}$ is the unperturbed direction from the observer to the source at the moment of observation and $h_{ij}$ is the metric perturbation due to the GW \citep{Pyne1996}.
It should be noted that this displacement corresponds to the component with respect to the tetrad of a cosmological static observer, i.e., a hypothetical observer at rest in the cosmological frame. In practice, the Gaia catalog contains processed measurements, where the Gaia RElativistic Model (GREM) reconstructs a single direction per observation \citep{Klioner2003, Crosta2010}. Hence, the formula above should be interpreted as the theoretical reference frame, while the actual Gaia measurements include the effects of the data processing and the scanning law.

\begin{figure*}
\includegraphics[width = 90mm]{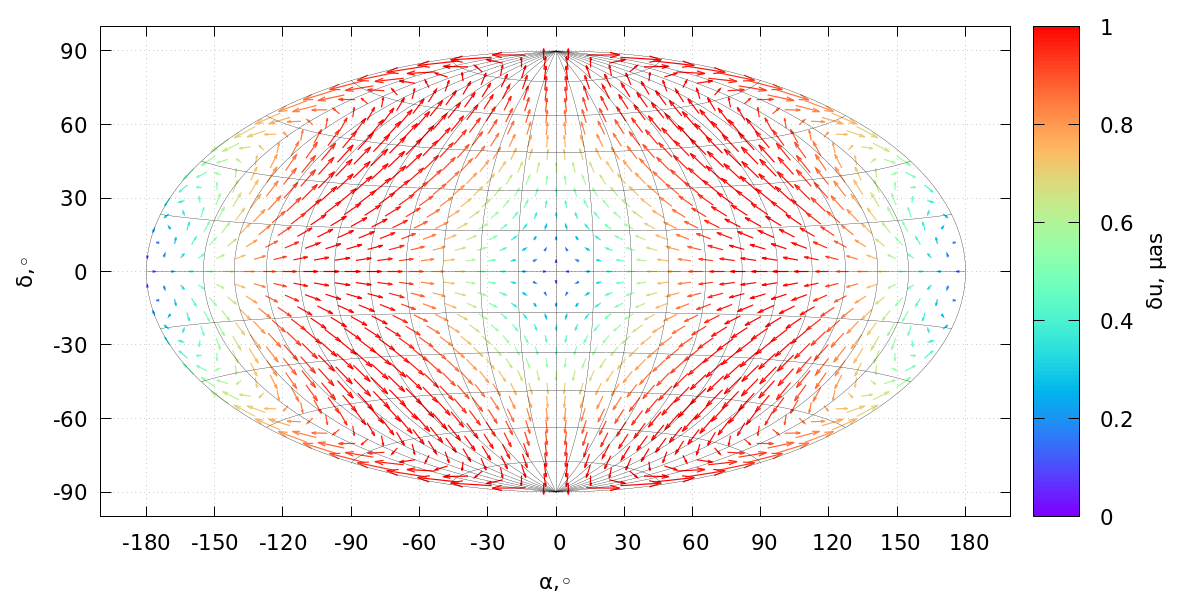}
\includegraphics[width = 90mm]{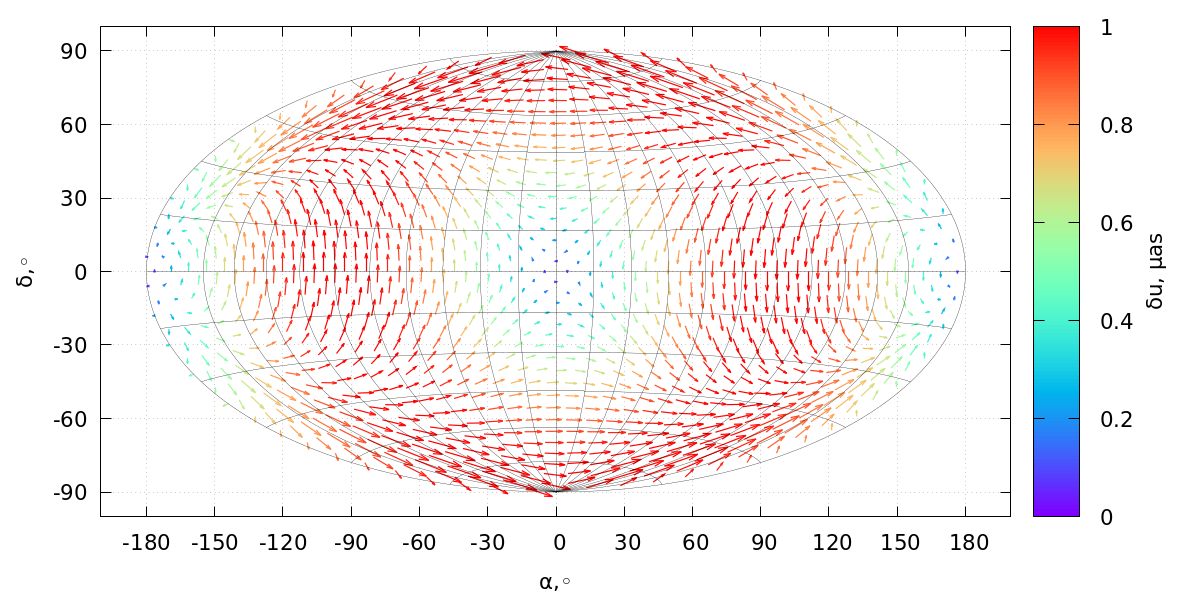}
\includegraphics[width = 90mm]{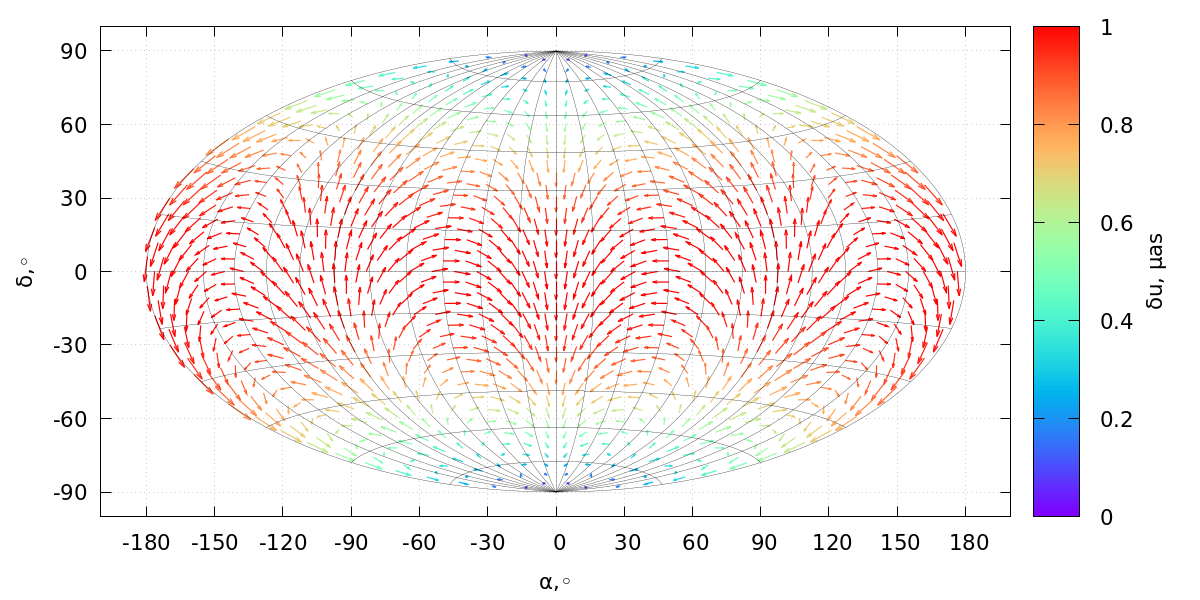}
\includegraphics[width = 90mm]{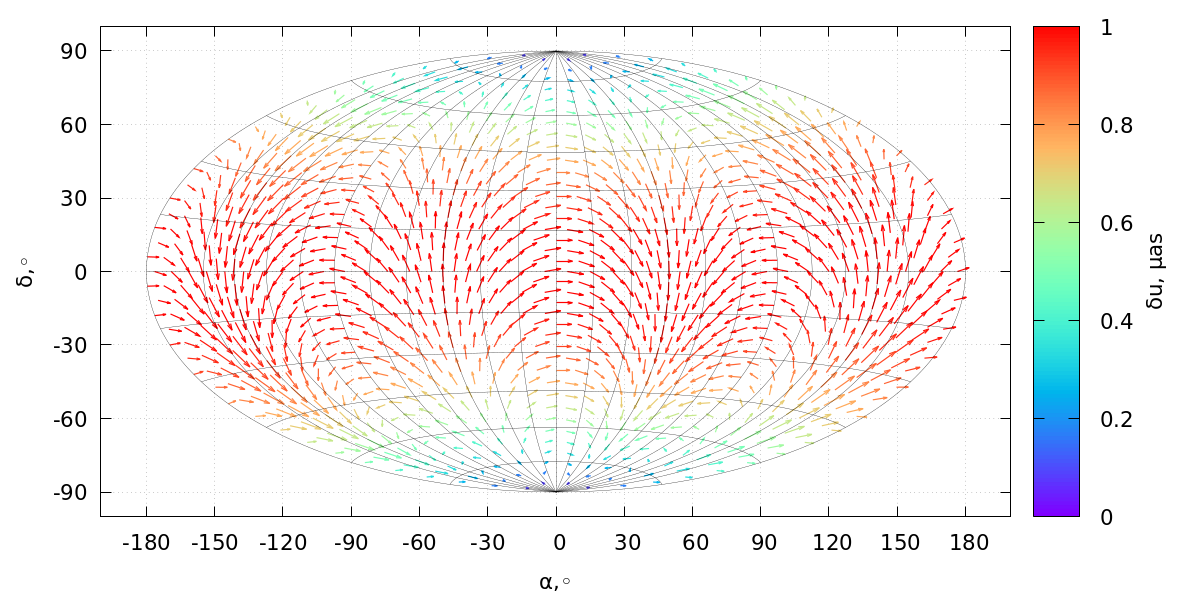}
\caption{Simulated apparent positional deflections $\delta \bf u$ of sources induced by a GW with strain amplitude $h=10^{-11}$, propagating toward the direction $\alpha=\delta=0^{\circ}$ (\textit{top panel}) and $\alpha= 0^{\circ}$ and $\delta=90^{\circ}$ (\textit{bottom panel}). The two GW polarizations are shown: "+" (\textit{left panel}) and "$\times$" (\textit{right panel}). The vector length and color encode the magnitude of the displacement $|\delta u|$. The maps use an Aitoff projection in equatorial coordinates.}
\label{fig:hc}
\end{figure*}

Right now, quasars are the most promising sources for GW detection with astrometry: they are extremely far extragalactic objects, meaning that their intrinsic proper motion should be way below the angular resolution scale of current optical telescope. Therefore, the observed proper motions must be caused by the imprint of GWs (either resolvable sources or a stochastic background) or other systematic effects. In this set-up, the actual observable for studying GW will be an angular proper motion, i.e., a displacement in the sky over a certain time interval $\Delta t$. Another key aspect is the frequency domain we can access with astrometric proper motion: these observables are the results of a linear fit from the astrometric solution and therefore they are a secular effect. It follows that, assuming that our sensitivity is strong enough for the detection and characterization of GWs, any effect on proper motions from frequencies higher than the inverse of the total observation time T, i.e. $1/T$, will be suppressed by the fitting pipeline. However, the lower limit on accessible frequencies is defined by the quasar lookback time.

It is clear that we are dealing with a different scenario from the standard radio astronomy GW search, since these experiments are designed to find periodic signals in a frequency range defined by their Nyquist limit and the total measuring time (roughly 20 years for PTA consortia at the time of writing). Different works have been proposed for the study of periodic GW signals from astrometric time series \citep {Book:2010pf, Klioner2018, Darling2018, Jaraba2023, Caliskan2024, Geyer2025}, but there are still difficult issues to overcome. 

Since quasars are extragalactic objects, their lookback time is cosmology dependent, while the time of data acquisition is 34 months for $Gaia$ DR3 \cite{Gaia:2021gsq} (corresponding to $\sim 11.3 \times 10^{-9}$ Hz) and will reach 10 years for $Gaia$ DR5 (corresponding to $\sim 3.2 \times 10^{-9}$ Hz). In this work, we adopt a $\Lambda$CDM cosmology with $\Omega_{M} = 0.315$, $\Omega_{\Lambda} = 0.685$, and $H_0 = 67 \ \text{km} \ \text{s}^{-1} \text{Mpc}^{-1}$ \cite{Planck:2018vyg}, from which follows a  lower limit of frequency of the order of $\sim10^{-18}$Hz.

In this study, we focus on the analysis of methods and their limitations in estimating the amplitude of GWs. We do not investigate the physical properties or nature of the GWs themselves but rather aim to quantify the statistical prospects for their detection in current astrometric data. To this end, we simulate astrometric deflections of quasar positions for a given direction, amplitude, and frequency of the GW source, using the formalism described by \cite{Geyer2025}. The resulting vector field of apparent positional deflections for the $h_+$ and $h_\times$ polarizations is illustrated in the figure \ref{fig:hc} below.

To extract GW signatures from astrometric data, two main analysis techniques are used.\\
(i) Hellings–Downs Correlation (HDC): This approach evaluates the angular correlations of proper motion pairs as a function of their separation, providing a statistical measure of the presence of a stochastic GW background \citep{HellingsDowns1983, Gwinn_1997, Darling_2025}.

(ii) Vector Spherical Harmonics (VSH): This method decomposes the vector field of proper motions on the celestial sphere into orthogonal multipoles, allowing the separation of GW-induced signals from other astrophysical and systematic effects. \citep{ Titov2011, Klioner2018, Jaraba2023, Geyer2025}.

Both methods were implemented within a highly parallelized data analysis pipeline, combined with GW modeling that accounts for the sky distribution of quasars and the uncertainties of their proper motions from the \textit{Gaia} DR3 catalog. Appropriate weighting schemes were applied to mitigate the impact of systematic effects in both methods.

The use of simulated astrometric deflections of quasar positions induced by a given GW allows us to assess the impact of the quasar sky distribution, the precision of proper motion measurements in the $Gaia$ DR3 catalog, and various data selection criteria on the estimation of the GW amplitude. This analysis is performed using both the HDC method and the VSH approach, allowing a comparative evaluation of their statistical behavior, sensitivity to systematics, and computational performance.

\subsection{Vector Spherical Harmonics}

Since astrometric measurements are defined on the celestial sphere, a natural approach to characterizing vector fields on such a manifold is provided by spherical harmonic functions. The application of VSH \citep{Arfken2005} to the analysis of vector fields defined on the unit sphere originates from the classical expansion of scalar fields in the basis of spherical harmonic functions  $Y_{\ell m}(\alpha, \delta)$:
\begin{align}
    \vec{Y}_{\ell m}^E &= \frac{1}{\sqrt{\ell(\ell + 1) }} \vec{\nabla} Y_{\ell m} (\alpha, \delta), \\
    \vec{Y}_{\ell m}^B &= -\frac{1}{\sqrt{\ell(\ell + 1) }} \mathbf{u} \times \vec{\nabla} Y_{\ell m} (\alpha, \delta),
\end{align}
where the superscripts $E$ and $B$ correspond to the spheroidal (electric-like) and toroidal (magnetic-like) modes, respectively. It follows that any vector field 
 $\vec{V} (\alpha, \delta)$ defined on the sphere can be expanded on the VSH basis as
\begin{equation}
    \vec{V} (\alpha, \delta) = \sum_{\ell = 1}^{\infty} \sum_{m = - \ell} ^\ell (s_{\ell m} \vec{Y}_{\ell m}^E (\alpha, \delta) + t_{\ell m} \vec{Y}_{\ell m}^B (\alpha, \delta))
\end{equation}
with $s_{\ell m}$ and $t_{\ell m}$ are complex expansion coefficients. Within this framework, VSH provide a powerful and flexible tool for studying the statistical and physical properties of quasar proper motions, as demonstrated in several previous works \cite{ Gwinn_1997, Vityazev_2009,  Titov2011, Mignard:2012xm, Klioner2021, Klioner2022}.

With VSH, we can already assess the impact of some systematic (and astrophysical) effects on astrometric solutions from the magnitude of the dipole contributions, i.e. the first order terms in the expansion. The term proportional to $\vec{Y}_{1 m}^E$ (spheroidal harmonic) is associated with the acceleration toward the Galactic center that the Solar System is experiencing, whereas the term $\vec{Y}_{1 m}^B$ (toroidal harmonic) is related to a rotation of the reference system (or a rotation of the universe). While the Galactic drift is an astrophysical effect that cannot be eliminated from astrometric measurements \citep{Klioner2021, Brown2025}, an ideal telescope should not exhibit any rotation of the reference frame (assuming that our universe is not rotating related to Mach’s principle \citep{Schiff}). 

Secular effects in quasar proper motions induced by passing GWs are expected to have their dominant contribution at quadrupole order $\ell = 2$. Assuming that such effects are caused by a stochastic gravitational-wave background (SGWB), one can relate the magnitude of total quadrupole power (from both \textit{E} and \textit{B} modes) to the energy density of the SGWB in the universe $\Omega_{GW}$ in the following way \citep{Darling2018}:
\begin{equation}
    \Omega_{GW} = \dfrac{6}{5} \dfrac{P_2}{4 \pi H_0^2},
\end{equation}
which in our adopted cosmology is equivalent to:
\begin{equation}
    \Omega_{GW} = 0.00042 \dfrac{P_2}{(1 \mu as / yr)^2} h_{70}^{-2} ,   
\end{equation}
where $P_2$ is the square of the total quadrupole power,
\begin{equation}
    P_2 = \sum_{m=-2}^{2} (s_{2m}^2 + t_{2m}^2).
\end{equation}
This equality is valid when the associated Legendre polynomials include the normalization factor $\sqrt{2}$, following the definition of \cite{Abramowitz1972}, as adopted in \cite{Vityazev_2009}. If a different normalization is used, all harmonics with $m\neq0$ should be multiplied by an additional factor of $\sqrt{2}$.

Although an SGWB can generate power at higher multipoles, the dominant contribution is expected in the quadrupolar ($\ell=2$) modes, as we show in Appendix \ref{App1}. In practice, the power at $\ell>2$ is often dominated by noise and systematic effects, and can therefore be used as a diagnostic of residual systematics in the catalog.

\subsection{Pair Correlations}

Another possible way to probing the effects of GWs in astrometric data is the analysis of proper motion correlations for all possible pairs of quasars. For such a method, first introduced in the context of PTA, one needs to define the angular correlation function (or HDC) \cite{HellingsDowns1983} between the GW induced perturbations on two different objects in the sky, which in this scenario comes down to quasar Since proper motions are two dimensional vector quantities, there is no unique way to define HDC for their individual components; however, a useful approach has been proposed \cite{Moore2017, Mihaylov2018, Darling_2025}.

To compute the proper motion components required for the astrometric HD analysis, it is necessary to determine the great-circle directions connecting each pair of quasars. In the equatorial coordinate system the components of the normal triad [ ${\bf p}$ ${\bf q}$ ${\bf r}$ ] are given by the matrix:
\begin{gather}
{\bf R} = \left( \begin{matrix}
    p_x & q_x & r_x \\
    p_y & q_y & r_y \\
    p_z & q_z & r_z  
  \end{matrix} \right) 
  =
 \left( \begin{matrix}
  -\sin \, \alpha & -\sin \,\delta \: \cos\, \alpha & \cos \, \delta \: \cos \, \alpha \\
  \cos \, \alpha & -\sin \, \delta \: \sin \, \alpha & \cos \, \delta \: \sin \, \alpha  \\
  0 & \cos \, \delta & \sin \, \delta  
  \end{matrix} \right) 
  \label{eq:n_matrix}
\end{gather}
For a sphere with radius equal to 1 the equatorial components of the proper motion $\hat{\boldsymbol{\mu}}_n$ in direction $\mathbf{\hat{n}}$ may thus be written in matrix form as:
\begin{gather}
\mathbf{\hat n}
=
\begin{pmatrix}
n_x \\ n_y \\ n_z
\end{pmatrix}
=
\mathbf R
\begin{pmatrix}
0 \\ 0 \\ 1
\end{pmatrix},
\qquad
\hat{\boldsymbol{\mu}}
=
\begin{pmatrix}
\mu_x \\ \mu_y \\ \mu_z
\end{pmatrix}
=
\mathbf R
\begin{pmatrix}
\mu^*_{\alpha} \\ \mu_\delta \\ 0
\end{pmatrix}
\end{gather}

In astrometric catalogs, the proper motion in right ascension is typically provided as $\mu_{\alpha}^* = \mu_{\alpha} \, \cos \delta$, where the factor $\cos \delta$ accounts for the convergence of meridians toward the celestial poles and converts the right ascension component into a true angular displacement on the celestial sphere.

\begin{figure}
    \centering
    \includegraphics[width=1.00\linewidth]{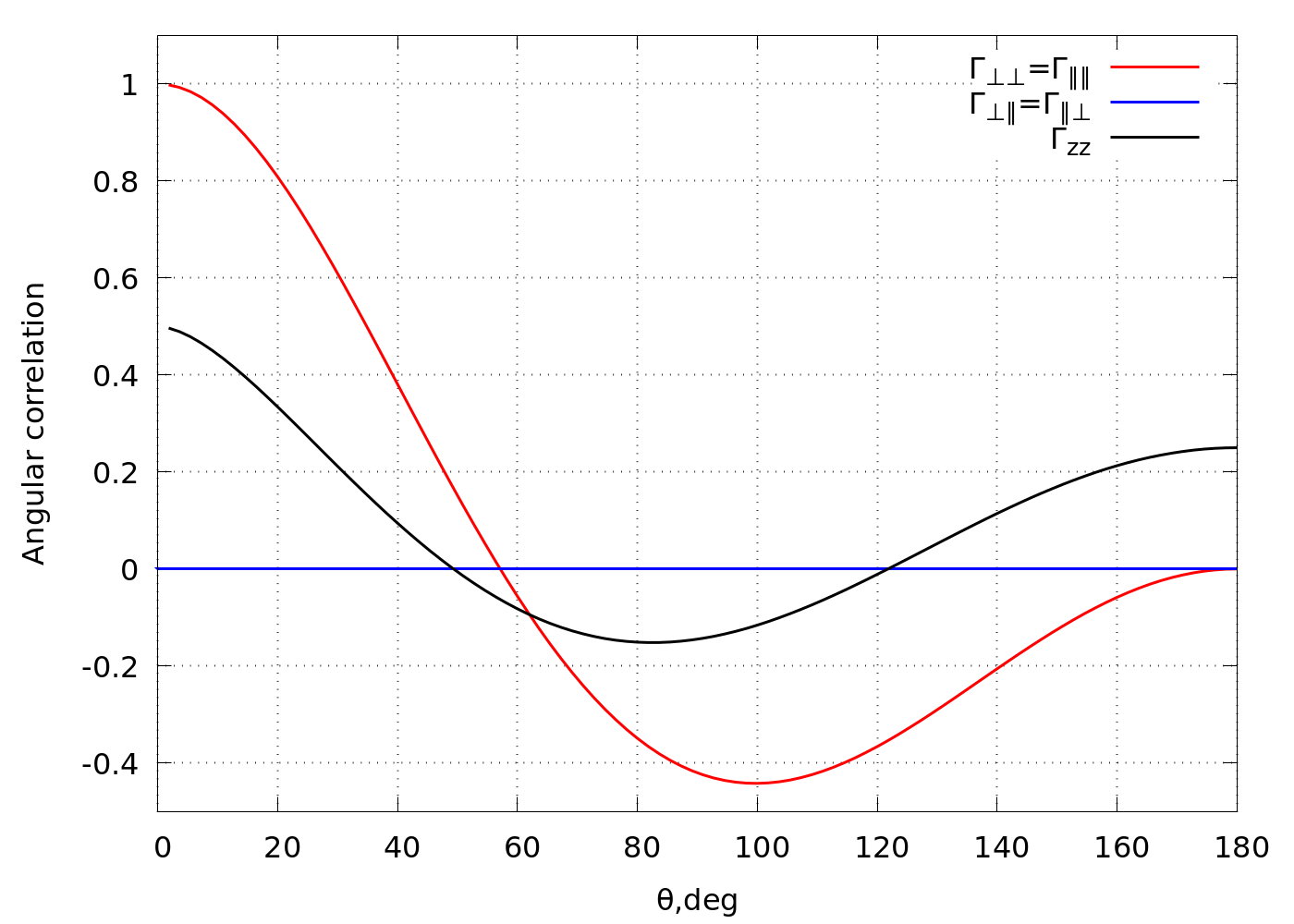}
    \caption{HDC for different GW induced displacements. In \textit{red} the parallel-parallel and perpendicular-perpendicular correlation function, in \textit{blue} the cross-terms (trivially zero for an isotropic GWB) and in \textit{black} the radial correlation function as measured by PTA (normalized to $\frac{1}{2}$ due to the non negligible star term for pulsars).}
    \label{fig:hd_curves}
\end{figure}

Two objects with proper motion vectors $\hat{\boldsymbol{\mu}}_n$ and $\hat{\boldsymbol{\mu}}_m$, located in the sky directions $\mathbf{\hat{n}}$ and $\mathbf{\hat{m}}$ are separated by an angular distance $\theta$, given by the dot product:
\begin{equation}
\cos \theta = \hat{\mathbf{n}} \cdot \hat{\mathbf{m}},
\end{equation}
where angle $\theta$ is the abscissa in the HDC in Fig. \ref{fig:hd_curves}.

For each pair of quasars, one needs to find the great circle on the sky connecting them, in order to decompose the proper motions in a component parallel to the great circle ($\hat{\mathbf{e}}_{\parallel,n}$ and $\hat{\mathbf{e}}_{\parallel,m}$) and a perpendicular one $\hat{\mathbf{e}}_{\perp}$:
\begin{equation}
\hat{\mathbf{e}}_{\perp} =
\frac{
\hat{\mathbf{n}} \times \hat{\mathbf{m}}}{\sqrt{1 - (\hat{\mathbf{n}} \cdot \hat{\mathbf{m}})^2}},  \ 
\hat{\mathbf{e}}_{\parallel,n} = \hat{\mathbf{e}}_{\perp} \times \hat{\mathbf{n}}, \
\hat{\mathbf{e}}_{\parallel,m} = \hat{\mathbf{e}}_{\perp} \times \hat{\mathbf{m}}.
\label{unit_vector}
\end{equation}

As a result, it is possible to define four different astrometric HDC: one describing the correlation between the parallel components of proper motions ($C_{\parallel \parallel}$), one for the perpendicular components($C_{\perp \perp}$), and two cross terms, which are not sensitive to an isotropic SGWB ($C_{\parallel \perp}$ and $C_{\perp \parallel}$):
\begin{equation}
\begin{split}
C_{\parallel \parallel}(\theta) = \left\langle (\boldsymbol{\hat\mu}_n \cdot \hat{\mathbf{e}}_{\parallel,n}) (\boldsymbol{\hat\mu}_m \cdot \hat{\mathbf{e}}_{\parallel,m})  \right\rangle_{\theta} , \\
C_{\perp\perp}(\theta) = \left\langle (\boldsymbol{\hat\mu}_n \cdot \hat{\mathbf{e}}_{\perp}) (\boldsymbol{\hat\mu}_m \cdot \hat{\mathbf{e}}_{\perp}) \right\rangle_{\theta}, \\
C_{\perp \parallel}(\theta) = \left\langle (\boldsymbol{\hat\mu}_n \cdot \hat{\mathbf{e}}_{\perp}) (\boldsymbol{\hat\mu}_m \cdot \hat{\mathbf{e}}_{\parallel,m}) \right\rangle_{\theta} ,\\
C_{\parallel \perp}(\theta) = \left\langle (\boldsymbol{\hat\mu}_n \cdot \hat{\mathbf{e}}_{\parallel,n}) (\boldsymbol{\hat\mu}_m \cdot \hat{\mathbf{e}}_{\perp})  \right\rangle_{\theta},
\label{eq:C_par_per}
\end{split}
\end{equation}
where the angular brackets denote an ensemble average over all source pairs with angular separation $\theta$ within a bin of width $\Delta \theta$. While these are the observed quantities, we must specify the corresponding theoretical expectation. Starting from the pairwise correlation of angular displacements in this decomposition, and approximating the SGWB as an unresolvable superposition of plane waves isotropically distributed over the sky, we get:
\begin{equation}
    \langle \delta u_{i,a} \ \delta u^*_{j,b} \rangle = \dfrac{1}{24} h_c^2(f) \Gamma_{ab} (\theta),
\end{equation}
where  $a,b \in \{||, \perp \}$, $h_c$ is the characteristic strain of the background and $\Gamma_{ab}$ is the proper angular correlation functions (the factor 1/24 comes from the normalization of $\Gamma_{ab}$ at $\theta=0$). As shown in \citep{Mihaylov2018}, there is an analytical solution for the angular correlation functions for an isotropic background:
\begin{equation}
\begin{split}
\Gamma_{|| \ ||} (\theta) = \Gamma_{\perp \perp} (\theta) =  1 - 7 \sin^2\left( \frac{\theta}{2}\right) \\ - 12 \sin^2\left( \frac{\theta}{2}\right) \, \tan^2\left( \frac{\theta}{2}\right) \, \ln \left[ \sin \left( \frac{\theta}{2}\right) \right] ,\\ 
\Gamma_{z \ z} (\theta) = \frac{1}{2}(1+\beta) + \frac{1}{4} \sin^2\left( \frac{\theta}{2}\right) + 3 \sin^2\left( \frac{\theta}{2}\right) \, \ln \left[ \sin \left( \frac{\theta}{2}\right) \right]. 
\end{split}
\label{eq:Gamma}
\end{equation}
where $\beta = 1$ for co-located pulsars and is zero otherwise and $\Gamma_{|| \perp} (\theta) = \Gamma_{\perp ||} (\theta)= 0$. These functions are plotted in Fig. \ref{fig:hd_curves}, together with the radial correlation function $\Gamma_{z\ z}$ measured by PTA experiments in \textit{black} color.

Additionally, since proper motions correspond to angular deflections measured over a time interval $\Delta t \sim 1/f$, the expected correlation (i.e., the observable quantity) acquires an additional factor of $f^2$. This arises because proper motions represent a secular accumulation of angular displacements, so the correlation scales with the square of the frequency of the underlying GW signal.
\begin{equation}
    \langle \mu_{i,a} \, \mu_{j,b} \rangle_\theta = \dfrac{1}{24} f^2 h_c^2(f) \Gamma_{ab} (\theta),
\end{equation}
or, alternatively
\begin{equation}
     \langle \mu_{i,a} \, \mu_{j,b} \rangle_\theta = \dfrac{1}{16 \pi^2}H_0^2 \Omega_{GW}(f) \Gamma_{ab} (\theta),
\end{equation}
where again $\Omega_{GW}$ is the energy density of GWs\citep{Darling2018}. \\
It is important to note that in the derivation above we assumed that only passing GWs (and white noise) contribute to the quasar proper motions. However, as discussed in the previous section, known systematic effects, such as the Galactic acceleration and the rotation of the reference frame, also affect measurements aimed at detecting GWs. Consequently, when performing a pair-correlation analysis using the proper motion from an astrometric catalog, these systematics may introduce spurious correlations. One way to mitigate this issue is to use VSH to estimate the dipole components of the proper motion and to correct the catalog by removing these contributions. After applying these corrections, the HD analysis can be performed on the modified catalog.

Although HDC and VSH analyses are often implemented as distinct computational approaches, they are formally related. The HDC formalism describes the two-point angular correlation function of the astrometric deflection field in angular configuration space, i.e. as a function of pair separation angle $\theta$, whereas the VSH decomposition represents the same vector field in harmonic space through a spherical harmonic expansion. In the statistically isotropic limit, the HDC function can be expressed in terms of the VSH power spectrum, with random phase information averaged. Such phase information is generally not relevant for a SGWB. Nevertheless, the two approaches emphasize different aspects of the signal and exhibit different sensitivity to observational systematics, incomplete sky coverage, and noise anisotropies.

\section{GW simulation}
\label{sec:simulation}

While in the section \ref{sec:methods} we presented a data analysis technique for a SGWB, in the case of a perfectly isotropic background, the analytical solution of the HDC function is the same, as we show in Appendix \ref{App1} by approximating the background as a superposition of a large number of independent GWs generated by a population of supermassive black hole binaries (SMBHBs). Therefore, in this scenario, we can effectively use a Single Source (SS) plane wave as mock signal, while still using the same formula presented above. Nevertheless, while the majority of the efforts are towards the detection of a stochastic background, we can still expect some SS signals in this frequency regime (although discussing the nature of the GW sources is beyond the scope of this article).

To investigate the astrometric signature of GWs on extragalactic sources, we simulate plane GWs with specified frequencies and propagation directions. As described in Section~\ref{sec:methods}, we investigate the influence of GWs on the proper motions of extragalactic objects by simulating a single plane GW with equal polarizations “$+$” and “$\times$”, characterized by a strain amplitude $h_+ = h_{\times} = 10^{-11}$, propagating in direction $\alpha = 45^{\circ}, \delta  = 45^{\circ}$. The GW frequency is chosen to be inversely proportional to the time baseline of the simulated proper motions, following \cite{Geyer2025}. 

This framework provides a clear geometric interpretation of the astrometric response: from the observer’s perspective, a GW induces a coherent and elliptical motion of all sources across the celestial sphere. The eccentricity of these ellipses is identical for all sources and is determined solely by the strain parameters of the wave, whereas their sizes and orientations also depend on the angular separation between the source direction and the GW propagation direction. A key outcome of this formulation is that the astrometric signature of a GW is, in some respects, analogous to that of astrometric binary systems. However, the essential distinction is that a GW produces a globally correlated signal, simultaneously affecting all sources on the sky.

In the limiting case of equal polarization amplitudes, the GW signal manifests itself as a synchronized circular motion of all sources on the sky, with an amplitude that depends only on the angle between the source direction and the direction of GW propagation. The maximum amplitude of the deviation on the sky occurs for a source located perpendicular to the GW propagation direction ($\theta =90^{\circ}$) whereas sources located close to the propagation axis ($\theta =0^{\circ}$ or $\theta =180^{\circ}$) experience only minimal deflections, as illustrated in Figure \ref{fig:hc}.

\subsection{Noise-free case}

\begin{figure*}
\includegraphics[width = 90mm]{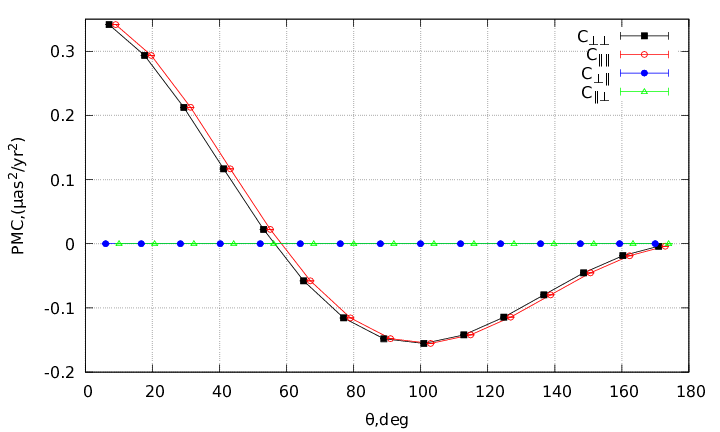}
\includegraphics[width = 90mm]{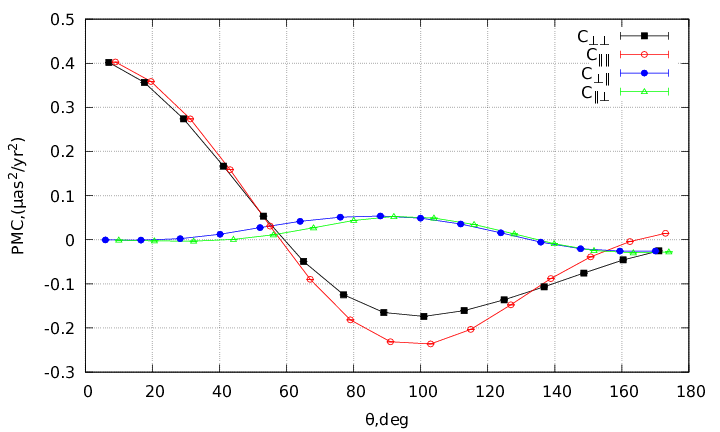}
\includegraphics[width = 90mm]{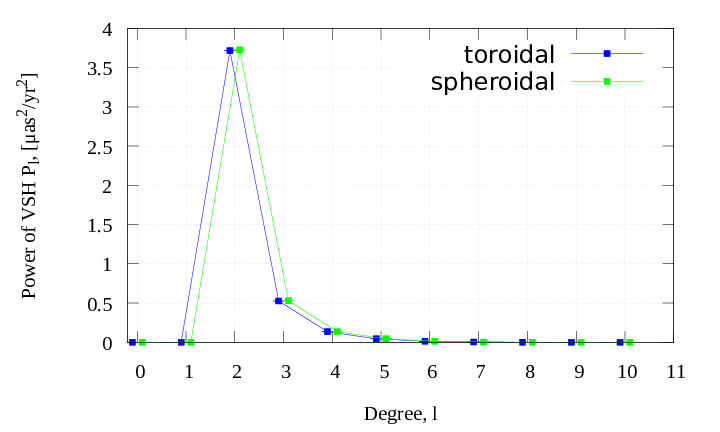}
\includegraphics[width = 90mm]{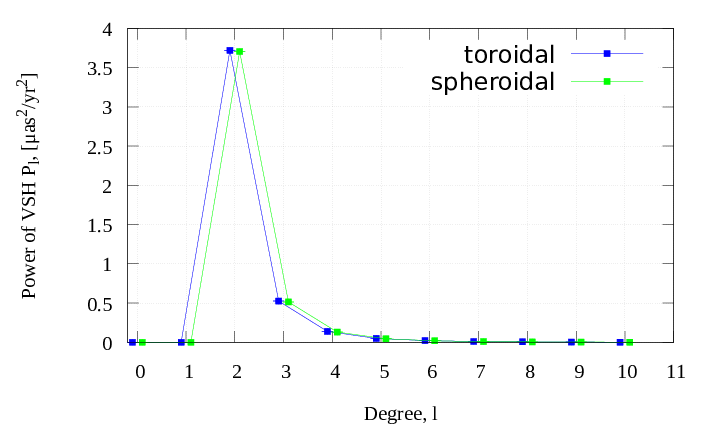}
\caption{ The \textit{top row} shows the HDC for all four PMC components, calculated from noise-free simulated data for a uniform sky distribution of sources (\textit{top left panels}) and for the actual distribution of 1.5 million quasars from the \textit{Gaia}-CRF3 catalog (\textit{top right panels}).
The \textit{bottom row} presents the power spectra of the VSH coefficients, $P_\ell$, for the same simulated datasets. For clarity, the toroidal (\textit{blue}) and spheroidal (\textit{green}) harmonic points are horizontally offset by $\Delta \ell = 0.1$.
The simulations assume a plane GW with equal “$+$” and “$\times$” polarizations, characterized by a strain amplitude $h_c = 10^{-11}$, propagating in the direction $\alpha = 45^{\circ}$, $\delta = 45^{\circ}$.}
\label{fig:NFM_model}
\end{figure*}

In Figure \ref{fig:NFM_model} (\textit{left panel}), we show the HD curves obtained from simulated proper motions of 30 000 uniformly distributed sources located at the centers of HEALPix pixels with \texttt{nside=50} \citep{Gorski2005}.
As seen in Figure \ref{fig:NFM_model} (\textit{top of the left panel}), in the noise-free case, the correlation components $C_{\perp \perp}$ and $C_{\parallel \parallel}$ are equal and match the theoretical HDCs (fig. \ref{fig:hd_curves}), while the cross terms, which are not sensitive to GWs ($C_{\parallel \perp}$ and $C_{\perp \parallel}$), remain zero for all angular separations $\theta$.

We performed a similar simulation of GW without adding noise, but using the actual positions of 1.5 million quasars from the $Gaia$-CRF3 catalog. As shown in Figure \ref{fig:NFM_model} (\textit{top of the right panel}), the correlation components $C_{\perp \perp}$ and $C_{\parallel \parallel}$ do not coincide for the angular separations $\theta > 40^\circ$ and deviate slightly from the theoretically expected HDC (fig. \ref{fig:hd_curves}).  

It is also noteworthy that the cross terms ($C_{\parallel \perp}$ and $C_{\perp \parallel}$), which are expected to be zero in the idealized uniform case, become non-zero starting from $\theta > 20^\circ$. They vary within certain ranges of $\theta$ and reach deviations of up to 25\% of the amplitudes of $C_{\perp \perp}$ and $C_{\parallel \parallel}$.  

This demonstrates that a non-uniform sky distribution of sources significantly affects the detectability of GWs and the accuracy of parameter estimation using the HDC method.

Figure \ref{fig:NFM_model} (\textit{bottom row}) presents the dependence of the VSH power, $P_l$, on the harmonic degree $l$, obtained from the simulated data. These graphs illustrate how the total power $P$ is distributed across harmonic degrees $\ell$, highlighting the quadrupole signal ($\ell = 2$) induced by a plane GW. As shown in the \textit{bottom left panel} of Fig. \ref{fig:NFM_model}, the powers of the toroidal and spheroidal harmonics coincide for all harmonic degrees, as expected for a GW induced astrometric signal.

GWs do not contribute to dipole harmonics ($\ell=1$), which is consistent with the quadrupole nature of GW. The dominant contribution appears at the quadrupole degree ($l=2$), where the measured power agrees with the theoretically expected value of $P_2^T = P_2^S \approx 3.7~\mu\mathrm{as}^2\,\mathrm{yr}^{-2}$. The power of higher-order harmonics decreases by approximately an order of magnitude with increasing $\ell$, indicating that the astrometric response is strongly dominated by the lowest non-vanishing multipole.

The physical origin of the dominance of the $\ell=2$ harmonics lies in the quadrupolar nature of GWs. A plane GW induces a coherent large-scale quadrupolar distortion of the apparent positions of sources on the celestial sphere, which is naturally captured by the $\ell=2$ VSH components. Thus, the values of $P_l$ directly reflect the relative amplitudes of the astrometric signal induced by a GW at different VSH degrees \citep{Klioner2018}.

As shown in the \textit{bottom right panel} of Fig. \ref{fig:NFM_model}, the non-uniform sky distribution of quasars in the $Gaia$-CRF3 catalog has a negligible impact on the estimated power of the VSH. This indicates that the VSH method is largely insensitive to inhomogeneities in the sky coverage of the sources. In contrast, the HDC method exhibits a strong dependence on the degree of uniformity of the source distribution.

This difference arises from the fact that the VSH formalism represents the global astrometric signal as an orthogonal expansion over the full sky, where the power at a given harmonic degree is obtained by integrating contributions from all directions. As a result, local variations in source density tend to average out and mainly contribute to statistical uncertainty rather than introducing a bias in the recovered harmonic power.

In contrast, the HDC method relies on pairwise correlations between sources at given angular separations. For a non-uniform sky distribution, certain angular bins become over- or under-sampled, leading to biased correlation estimates and increased leakage between correlation components. Consequently, the HDC approach is significantly more sensitive to sky inhomogeneities than the VSH-based analysis.

\subsection{Noise case}
\label{subsec:Noise}

One of the primary goals of this work is to investigate how non-uniform \textit{Gaia} DR3 quasar distributions and inhomogeneous proper motion uncertainties, together with their correlations, affect the detectability of GWs. To investigate the impact of observational uncertainties on the detection of GWs, we perform simulations including Gaussian noise in the quasar proper motions. The noise is modeled using the covariance matrices of the \textit{Gaia}-CRF3 proper motions, taking into account both the standard errors in proper motion uncertainties and their correlation.
\begin{gather}
\boldsymbol{Cov}
=
\begin{pmatrix}
\sigma^2_{\mu_{\alpha}} & \rho \sigma_{\mu_{\alpha}}\sigma_{\mu_{\delta}} & 0\\ 
\rho \sigma_{\mu_{\alpha}}\sigma_{\mu_{\delta}}  &  \sigma^2_{\mu_{\delta}} & 0\\
 0 & 0 & 0
\end{pmatrix}
\label{eq:Covmatrix}
\end{gather}
where $\sigma_{\mu_{\alpha}}$ and $\sigma_{\mu_{\delta}}$ are the standard errors of the proper motions in right ascension and declination, respectively, $\rho$ is the correlation coefficient between the proper motion components in right ascension and declination.

This noise is scaled according to the proper motion uncertainties of the 1.5 million quasars in the $Gaia$-CRF3 catalog (selected using criteria described in Sect. \ref{res1}) using a lower-triangular Cholesky matrix $\mathbf{L L^T = Cov }$ and defined as:
\begin{equation}
\mathbf{L} =
\begin{pmatrix}
\sigma_{\mu_\alpha} & 0  & 0 \\
\rho \, \sigma_{\mu_\delta} & \sigma_{\mu_\delta}  \, \sqrt {1-\rho^2}  & 0 \\
 0  & 0  & 0 
\end{pmatrix}
\label{eq:Cholesky}
\end{equation}

Then, the contribution of Gaussian noise in the modeling of proper motions can be represented as additional components $\delta\mu_\alpha$ and $\delta\mu_\delta$, which are determined using the formula:
\begin{equation}
\begin{pmatrix}
\delta \mu_\alpha \\
\delta \mu_\delta \\
0
\end{pmatrix}
=
\mathbf{L} 
\begin{pmatrix}
g_1 \\
g_2 \\
0
\end{pmatrix}
\end{equation}
where $g_1$ and $g_2$ are independent Gaussian random variables with $\sigma=1$. This procedure ensures correlated Gaussian perturbations consistent with the observed uncertainties and correlation of quasars proper motions in \textit{Gaia}-CRF3.

Our main assumption is that the simulated quasar proper motions contain only the distortions induced by the modeled GW, combined with Gaussian noise. Although this represents an idealized scenario, it allows us to isolate the GW signal and to systematically evaluate the performance of the analysis methods considered in this work.

An accurate determination of the HDC curve requires a correct weighting of the measured proper motions. This weighting must account for both the uncertainties of the proper motion components and their correlation coefficients. Under these conditions, the components of the covariance matrix in the cartesian equatorial coordinate system can be expressed as follows: $ \boldsymbol{\hat{C}ov} = \mathbf R \boldsymbol{Cov} \mathbf R^{\mathrm{T}}$, 
where $\mathbf R^{\mathrm{T}}$ denotes the transpose of the normal triad matrix defined in \eqref{eq:n_matrix}.
To correctly weight the proper motions when computing the correlation curve, it is necessary to obtain the scalar projection of the covariance matrix $\boldsymbol{\hat{C}ov}$ along the direction parallel to the great circle $\hat{\mathbf{e}}_{\parallel}$ defined by two objects with sky directions $\hat{\mathbf{n}}$ and $\hat{\mathbf{m}}$, as well as along the perpendicular direction $\hat{\mathbf{e}}_{\perp}$:
\begin{equation}
\begin{split}
\sigma^2_{\parallel,n} = \mathbf{\hat{e}_{\parallel,n}}^\top \cdot \boldsymbol{\hat{C}ov_{n}} \cdot \mathbf{\hat{e}_{\parallel,n}}, \, 
\quad
\sigma^2_{\perp,n} = \mathbf{\hat{e}_{\perp}}^\top \cdot \boldsymbol{\hat{C}ov_{n}} \cdot 
\mathbf{\hat{e}_{\perp}}, \\
\quad
\sigma^2_{\parallel,m} = \mathbf{\hat{e}_{\parallel,m}}^\top \cdot \boldsymbol{\hat{C}ov_{m}} \cdot \mathbf{\hat{e}_{\parallel,m}}, \, 
\quad
\sigma^2_{\perp,m} = \mathbf{\hat{e}_{\perp}}^\top \cdot \boldsymbol{\hat{C}ov_{m}} \cdot \mathbf{\hat{e}_{\perp}}.
\quad
\end{split}
\label{eq:sig}
\end{equation}
where $\mathbf{\hat{e}}^\top$ denotes the transpose of the unit vector corresponding to the great-circle direction for a given object, defined using Eqs.~\ref{unit_vector}. The total variance of the proper motion uncertainties projected onto these directions for a pair of objects is then given by the following expressions:
\begin{equation}
\begin{split}
W_{\parallel\parallel} =  (\boldsymbol{\hat\mu}_n \cdot \hat{\mathbf{e}}_{\parallel,n})^2 \, \sigma^2_{\parallel,m} + (\boldsymbol{\hat\mu}_m \cdot \hat{\mathbf{e}}_{\parallel,m})^2 \,  \sigma^2_{\parallel,n} + \sigma^2_{\parallel,n} \, \sigma^2_{\parallel,m}, \\
\quad W_{\perp\perp} =  (\boldsymbol{\hat\mu}_n \cdot \hat{\mathbf{e}}_{\perp})^2 \, \sigma^2_{\perp,m} + (\boldsymbol{\hat\mu}_m \cdot \hat{\mathbf{e}}_{\perp})^2 \, \sigma^2_{\perp,n} + \sigma^2_{\perp,n} \, \sigma^2_{\perp,m},\\
W_{\perp\parallel} =  (\boldsymbol{\hat\mu}_n \cdot \hat{\mathbf{e}}_{\perp})^2 \, \sigma^2_{\parallel,m} + (\boldsymbol{\hat\mu}_m \cdot \hat{\mathbf{e}}_{\parallel,m})^2 \, \sigma^2_{\perp,n} +  \sigma^2_{\perp,n}\, \sigma^2_{\parallel,m}, \\
W_{\parallel\perp} =  (\boldsymbol{\hat\mu}_n \cdot \hat{\mathbf{e}}_{\parallel,n})^2 \, \sigma^2_{\perp,m} + (\boldsymbol{\hat\mu}_m \cdot \hat{\mathbf{e}}_{\perp})^2 \, \sigma^2_{\parallel,n} +  \sigma^2_{\parallel,n} \, \sigma^2_{\perp,m}.
\end{split}
\label{eq:varianc}
\end{equation}
Finally, these variances are used to weight the proper motion correlations $C_{ab}(\theta)$ for all possible combinations:
\begin{equation}
\begin{split}
C_{\parallel \parallel}(\theta) = \left\langle (\boldsymbol{\hat\mu}_n \cdot \hat{\mathbf{e}}_{\parallel,n}) (\boldsymbol{\hat\mu}_m \cdot \hat{\mathbf{e}}_{\parallel,m}) / W_{\parallel \parallel} \right\rangle_{\theta} / \left\langle W_{\parallel \parallel} \right\rangle_{\theta}, \\
C_{\perp\perp}(\theta) = \left\langle (\boldsymbol{\hat\mu}_n \cdot \hat{\mathbf{e}}_{\perp}) (\boldsymbol{\hat\mu}_m \cdot \hat{\mathbf{e}}_{\perp})/ W_{\perp\perp}  \right\rangle_{\theta}, / \left\langle W_{\perp\perp} \right\rangle_{\theta}, \\
C_{\perp \parallel}(\theta) = \left\langle (\boldsymbol{\hat\mu}_n \cdot \hat{\mathbf{e}}_{\perp}) (\boldsymbol{\hat\mu}_m \cdot \hat{\mathbf{e}}_{\parallel,m}) / W_{\perp\parallel} \right\rangle_{\theta} / \left\langle W_{\perp \parallel} \right\rangle_{\theta},\\
C_{\parallel \perp}(\theta) = \left\langle (\boldsymbol{\hat\mu}_n \cdot \hat{\mathbf{e}}_{\parallel,n}) (\boldsymbol{\hat\mu}_m \cdot \hat{\mathbf{e}}_{\perp}) / W_{\parallel \perp} \right\rangle_{\theta} / \left\langle W_{\parallel \perp} \right\rangle_{\theta}
\label{eq:C_par_per_w}
\end{split}
\end{equation}
In this approach, the same quasar can be assigned different effective weights depending on the specific source pair and on the projected component of the proper motion. As shown in the appendix \ref{App2:weighting}, this leads to a significant improvement in the robustness and precision of the recovered HDC amplitudes while providing a statistically consistent treatment of correlated astrometric uncertainties. 

In realistic astrometric catalogs such as \textit{Gaia}-CRF3, the uncertainties of the proper motion components are heteroscedastic, and the errors in $\mu^*_{\alpha}$ and $\mu_\delta$ are generally correlated. Because the observation equations have different uncertainties and include correlated errors, the statistically optimal solution is obtained using the generalized least-squares (GLS) method.  

Stacking the equations for all sources, the GLS estimator of the vector of VSH coefficients $\boldsymbol{x}$ is given by:
\begin{equation}
\boldsymbol{x} = 
\left(\boldsymbol{A}^{\rm T} \boldsymbol{Cov}^{-1} \boldsymbol{A}\right)^{-1} 
\boldsymbol{A}^{\rm T} \boldsymbol{Cov}^{-1} \boldsymbol{\mu},
\end{equation}
where $\boldsymbol{A}$ is the design matrix $2N \times m$ contains the values of the $m$ VSH basis functions evaluated at the positions of all $N$ sources, $\boldsymbol{Cov}^{-1}$ is the inverse of the full covariance matrix of proper motions (see Eq.~\ref{eq:Covmatrix}), and $\boldsymbol{\mu}$ is the $2N$-dimensional proper motion vector. The number of equations is twice the number of sources because, for each source, two equations are included: one for $\mu_{\alpha*}$ and one for $\mu_\delta$.

Using the Cholesky decomposition of each source covariance (see Eq.~\ref{eq:Cholesky}), and multiplying the observation equations by $\boldsymbol{L}^{-1}$ yields the transformed system of equations:
\begin{equation}
\tilde{\boldsymbol{\mu}} = \tilde{\boldsymbol{A}} \boldsymbol{x} + \tilde{\boldsymbol{\varepsilon}},
\end{equation}
with
\begin{equation}
\tilde{\boldsymbol{\mu}} = \boldsymbol{L}^{-1} \boldsymbol{\mu}, 
\qquad
\tilde{\boldsymbol{A}} = \boldsymbol{L}^{-1} \boldsymbol{A}.
\end{equation}
where $\tilde{\boldsymbol{\varepsilon}}$ is the “weighted and de-correlated noise” after accounting for the correlations and the different uncertainties between $\mu^*_{\alpha}$ and $\mu_\delta$.
This transformation effectively standardizes the scale of the errors and de-correlates them. The problem then reduces to an ordinary least-squares (OLS) solution applied to these transformed equations.

Appendix \ref{App2:weighting} presents a comparison of the HDC and VSH methods to estimate the GW amplitude using the same simulated data set, highlighting the differences between analyzes performed with and without weighting schemes.

\begin{figure*}
\includegraphics[width = 90mm]{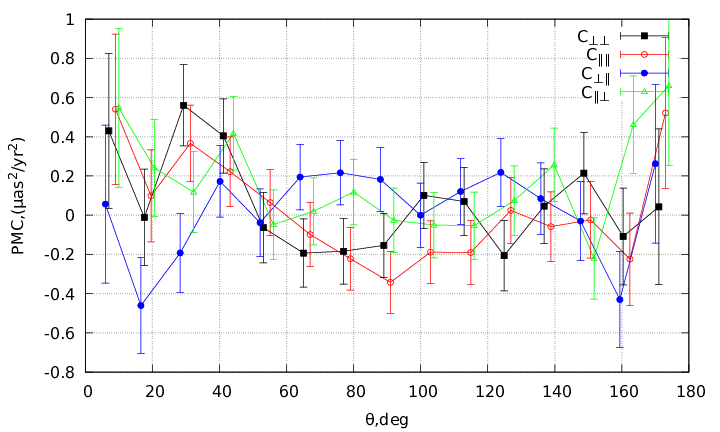}
\includegraphics[width = 90mm]{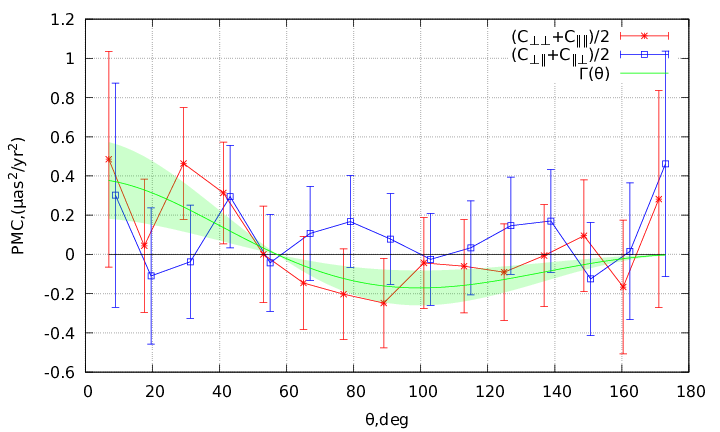}
\includegraphics[width = 90mm]{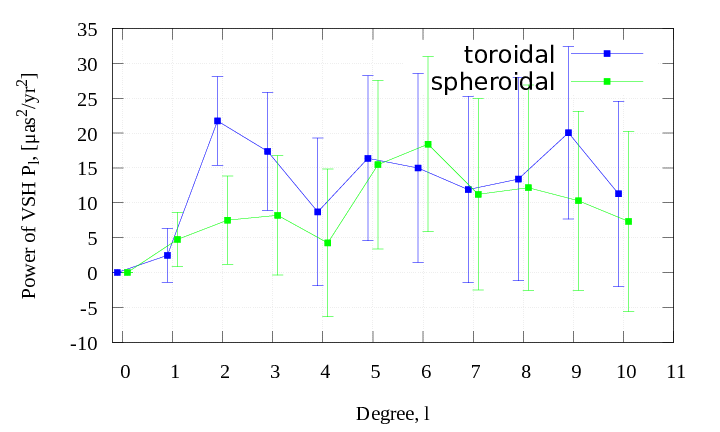}
\includegraphics[width = 90mm]{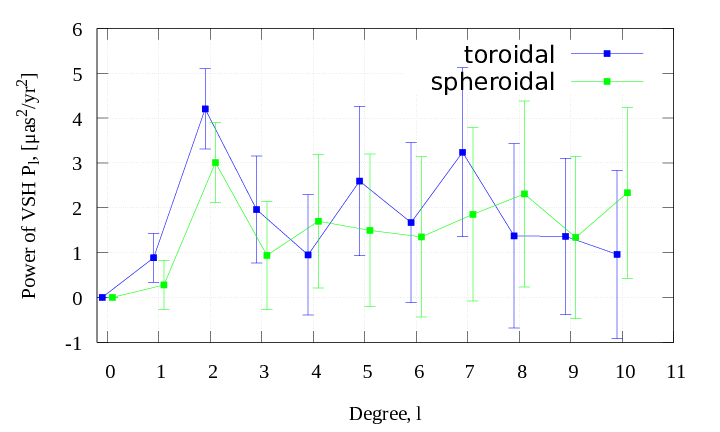}
\caption{The \textit{top row} shows the HDC for all four PMC components, calculated from simulated proper motions for the actual quasar sky distribution, with \textit{Gaia}-CRF3 error-dependent noise added (\textit{top left panel}). The \textit{top right panel} presents the averaged correlation components $C_{\perp\perp}$ and $C_{\parallel\parallel}$ (\textit{red}),  together with the cross terms ($C_{\parallel\perp}$ and $C_{\perp\parallel}$) (\textit{blue}). Also shown are the fitted HDC function  $\Gamma(\theta)$ and its corresponding $95\%$ confidence interval (\textit{green}), derived from these correlation curves.
The \textit{bottom row} shows the power spectra of the VSH coefficients $P_l$ for the same simulated dataset (\textit{bottom left panel}), and for a simulation in which the quasar proper motion uncertainties are reduced by a factor of three, as expected for future \textit{Gaia} DR4 (\textit{bottom right panel}). For clarity, the toroidal (\textit{blue}) and spheroidal (\textit{green}) harmonic points are horizontally offset by 0.1 in $\ell$. The simulations assume a plane GW with equal “$+$” and “$\times$” polarizations, characterized by a strain amplitude $h_c = 10^{-11}$, propagating in the direction $\alpha = 45^{\circ}, \delta = 45^{\circ}$.}
\label{fig:GN_model}
\end{figure*}

Figure~\ref{fig:GN_model} (\textit{top left panel}) presents the HD curves derived from GW simulations that include noise, using the actual \textit{Gaia}-CRF3 quasar positions together with their proper motion uncertainties and correlations. 
The inclusion of noise further amplifies the deviations of the components $C_{\perp\perp}$ and $C_{\parallel\parallel}$ from the theoretical HD curve and introduces additional scatter. This clearly highlights the combined impact of uneven sky coverage and non-uniform proper motion uncertainties on the detectability of GWs and on the accuracy of parameter estimation using the HDC method. The cross terms ($C_{\parallel\perp}$ and $C_{\perp\parallel}$) depart from zero over the entire range of angular separations $\theta$, demonstrating that uncertainties in quasar proper motions contribute significantly to the observed correlations, in addition to the effects of their non-uniform distribution on the celestial sphere.  

Despite the substantial scatter of the correlation components shown in Fig. \ref{fig:GN_model} (\textit{top left panel}), we performed a fit of the correlation components using the functional form given by Eq. \ref{eq:Gamma}. This yields a value of $0.4238 \pm 0.1281~\mu\mathrm{as}^2 \mathrm{yr}^{-2}$ (fig. \ref{fig:GN_model} \textit{top right panel}), which is very close to the model expectation of $0.3533~\mu\mathrm{as}^2 \mathrm{yr}^{-2}$. The 95\% confidence interval shown in this figure  \textit{in green} corresponds to the uncertainty of the fitted amplitude parameter of the HDC function.

The \textit{ bottom left panel} of Fig.~\ref{fig:GN_model} shows the VSH power as a function of the harmonic degree $\ell$, calculated from simulated proper motions for the actual quasar sky distribution, with \textit{Gaia}-CRF3 error-dependent noise added. As can be seen, the power of the $\ell=2$ harmonics significantly exceeds the expected values that were introduced into the proper motions through the simulated GW. Higher-order harmonics also appear, resulting from the superposition of several factors: significant noise in the proper motions and its dependence on the source positions. It can be concluded that the non-uniform sky distribution of sources, combined with substantial proper motion uncertainties and the presence of correlations, leads to a spurious increase in the inferred GW amplitude by a factor of 3–5 when using the VSH method.  

However, if the random component of proper motions is reduced by a factor of three, i.e., using $0.3\,\sigma_{\mu_\alpha}$ and $0.3\,\sigma_{\mu_\delta}$ in Eq.~\ref{eq:Cholesky}, as expected in the next \textit{Gaia} data release, the VSH method allows the successful detection of GWs with a strain amplitude $h_c = 10^{-11}$, as shown in the \textit{bottom right panel} of Fig.~\ref{fig:GN_model}.

As seen in the \textit{bottom right panel} of Fig.~\ref{fig:GN_model}, the power of the second-order harmonics for the toroidal (B-mode) and spheroidal (E-mode) components are similar and consistent with the expected values within their respective uncertainties. Almost all higher-order harmonic powers do not exceed their measurement errors, and their amplitudes are roughly an order of magnitude smaller than in the \textit{bottom left panel}, confirming the threefold reduction in proper motion uncertainties.

Thus, we conclude that the VSH method is more statistically robust, less sensitive to uneven object sampling, more sensitive to noise in GW detection, and computationally faster. The computational complexity of the VSH method scales linearly with the number of sources $N$. In contrast, the HDC method exhibits a higher sensitivity to injected GW like angular correlations under idealized conditions, but its complexity scales as $N^2$ and is more prone to cherry picking. Therefore, applying the HDC method to astrometric global iterative solution (AGIS) residuals or to catalogs containing tens of millions of sources requires appropriate algorithmic modifications and the use of parallel computations on GPUs.

This result suggests that, given the statistical precision of the proper motions for the 1.5 million quasars in the \textit{Gaia}-CRF3 catalog, the HDC and VSH methods could detect GWs with a strain amplitude of $h_c \gtrsim 10^{-11}$ with an empirical signal-to-noise ratio of approximately 3. The statistical significance of the recovered amplitudes is quantified through the ratios ($A_{\rm HDC}/\sigma_A$) and ($P_2/\sigma_{P_2}$), which serve as empirical signal-to-noise estimators for HDC and VSH analyzes, respectively. Appendix~\ref{App3} presents the results of 1000 Monte Carlo injection-and-recovery simulations of the GW strain amplitude for both the HDC and VSH methods, including the median recovered amplitudes, the central 15th–85th percentile ranges, the 95\% confidence intervals, and the fractions of realizations with ($h_{rec} > 2\sigma$) and ($h_{rec} > 3\sigma$).

However, this conclusion holds only under idealized assumptions, namely: (i) the absence of systematic errors in the \textit{Gaia} catalog and (ii) a regime in which a portion of the GW-induced elliptical motion can be approximated by a linear trend corresponding to a constant proper motion. This occurs when the GW period is approximately twice the observation time, $P_{\mathrm{GW}} \approx 2\,T_{\mathrm{obs}}$ \citep{Gwinn_1997}. In the opposite limits, the signal is largely absorbed by the AGIS astrometric solution ($P_{\mathrm{GW}} \ll 2\,T_{\mathrm{obs}}$) or varies too slowly to produce a measurable proper motion signature ($P_{\mathrm{GW}} \gg 2\,T_{\mathrm{obs}}$).

In reality, the presence of systematic errors in the \textit{Gaia} catalog, as well as a mismatch between the observational time span and the GW period, can significantly suppress the GW signature in quasar proper motions, as demonstrated by \citep{Geyer2025}. Consequently, under realistic conditions, the upper limit on the detectable GW strain amplitude using the HDC or VSH methods is expected to be higher by a factor of approximately $2-3$ using data from \textit{Gaia}-CRF3 quasars.

\section{Results}\label{sec:results}

\subsection{Gaia-CRF3 catalog analysis}
\label{res1}

 We apply both the HDC method and the VSH approach to the \textit{Gaia}-CRF3 catalog \citep{Klioner2022}. In practice, the true proper motions of quasars are expected to be essentially zero. However, due to the limited astrometric precision of the \textit{Gaia} measurements, the observed proper motions are non-zero and dominated by measurement noise. The systematic contribution to the proper motions caused by the acceleration of the solar system barycenter and by GWs (and also, possibly, the presence of systematic errors) is at the level of a few $\mu{\rm as~yr^{-1}}$, which is several orders of magnitude smaller than the typical proper motion uncertainties.

For each quasar, an effective signal-to-noise ratio can be defined as
\begin{equation}
(\mathrm{S/N})^2 = \frac{1}{1-\rho^2} \left[
\frac{\mu_{\alpha}^2}{\sigma_{\mu_{\alpha}}^2} + \frac{\mu_{\delta}^2}{\sigma_{\mu_{\delta}}^2}
- 2 \rho \frac{\mu_{\alpha}\mu_{\delta}}{\sigma_{\mu_{\alpha}}\sigma_{\mu_{\delta}}}
\right],
\end{equation}
A moderate threshold of $\mathrm{S/N}<3$ is applied to reject sources with high proper motions while retaining a sufficiently large number of quasars. In addition, we apply a quality cut of $\mathrm{RUWE}<1.2$ (Renormalized Unit Weight Error), which is a reliable goodness of fit statistic selecting sources with well behaved astrometric solutions. This combined selection balances the trade-off between reducing measurement noise and maintaining a sample size large enough to reliably detect the correlated signal induced by GWs. As a result, we obtain a final sample comprising positions, proper motions, their uncertainties, and correlations for 1 \, 505 \, 427 quasars. The statistical properties of the selected quasar sample from the \textit{Gaia}-CRF3 catalog are presented in Table \ref{tab:quasar_stats}. 

\begin{table*}
\centering
\caption{Statistical properties of the selected sample of 1\,505\,427 quasars from the \textit{Gaia}-CRF3 catalog.}
\label{tab:quasar_stats}
\begin{tabular}{lccccccc}
\hline\hline
Parameter & Mean & SD & Minimum & Maximum & Median & Skewness & Kurtosis \\
\hline
$\mu^*_\alpha$ ($\mu$as~yr$^{-1}$) & $-0.68$ & 793     & $-8\,521$ & 8\,728 & $3.26\times10^{-2}$ & $-0.008$ & 6.609\\
$\mu_\delta$ ($\mu$as~yr$^{-1}$) & $-1.37$ & 736 & $-8\,566$ & 9\,297 & $-1.23\times10^{-2}$ & $-0.010$ & 6.839 \\
$\sigma_{\mu^*_\alpha}$ ($\mu$as~yr$^{-1}$) & 653 & 476 & 9.72   & 3\,365 & 516     & 1.583 & 2.872\\
$\sigma_{\mu_\delta}$ ($\mu$as~yr$^{-1}$) & 606 & 441 & 12.98  & 3\,387 & 478     & 1.670 & 3.396\\
\hline
\end{tabular}
\end{table*}

As a first step in the analysis of the proper motions of the selected quasar sample, we applied the VSH formalism to estimate and remove large-scale systematic effects, represented by the lowest-order harmonics. These terms account for the residual rotation of the celestial reference frame and for the apparent dipole pattern induced by the acceleration of the Solar System barycenter.

\begin{figure*}
\includegraphics[width = 90mm]{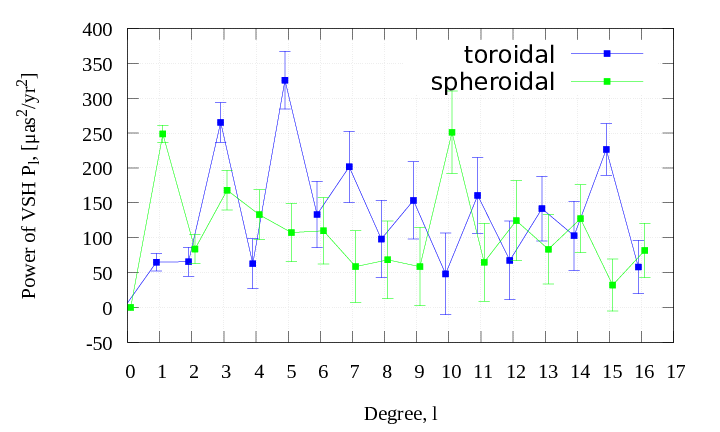}
\includegraphics[width = 90mm]{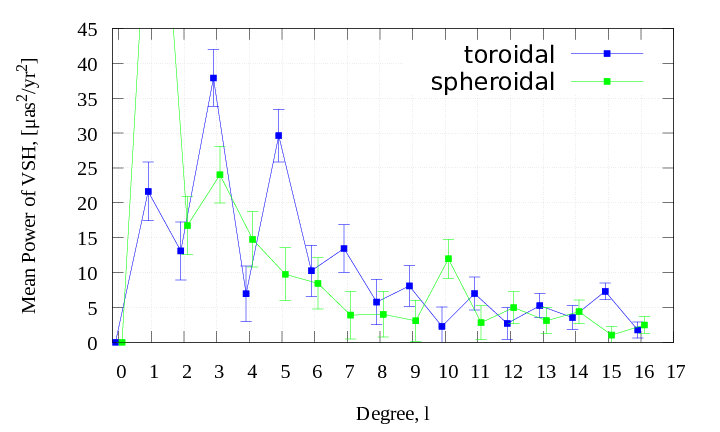}
\includegraphics[width = 90mm]{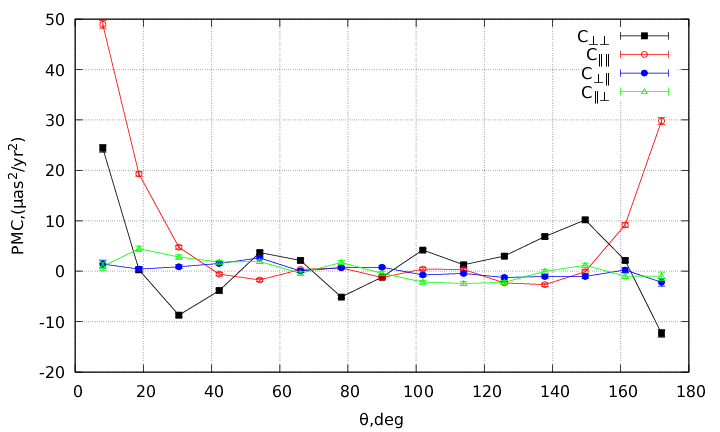}
    \includegraphics[width = 90mm]{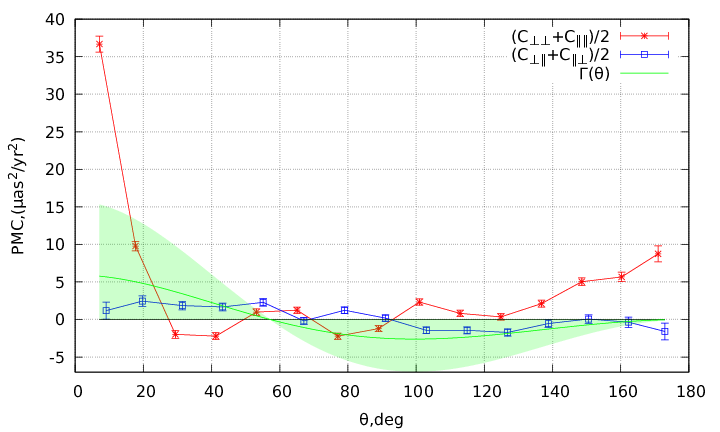}
\caption{The \textit{top row} shows the power spectra of the VSH coefficients (\textit{left panel}) and the mean toroidal and spheroidal powers (\textit{right panel}) characterizing systematics in the proper motions of $\sim$1.5 million quasars from the \textit{Gaia}-CRF3 catalog. For clarity, the toroidal (\textit{blue}) and spheroidal (\textit{green}) harmonic coefficients are horizontally offset by $\Delta\ell = 0.1$.
The \textit{bottom row} displays the correlation components $C_{\perp\perp}$ and $C_{\parallel\parallel}$ together with the cross terms $C_{\parallel\perp}$ and $C_{\perp\parallel}$ (\textit{left panel}). Also the \textit{right panel} shows the averaged correlation components and the best-fitting HDC function $\Gamma(\theta)$ with its corresponding $95\%$ confidence interval (\textit{green}), obtained after applying a dipole-mode correction based on the proper motions of $\sim$1.5 million \textit{Gaia}-CRF3 quasars.}
\label{fig:gaiacrf3}
\end{figure*}

Figure \ref{fig:gaiacrf3} (\textit{top left panel}) presents the power of the VSH coefficients as a function of harmonic degree, while the \textit{top right panel} shows the corresponding mean power of each toroidal or spheroidal harmonic using formula $\tilde{P}_l=P_l/(2l+1)$. The mean power and the corresponding uncertainties derived from our sample are in good agreement with the results reported by \citep{Klioner2022}. It is also worth noting that the powers of higher-order harmonics derived from the actual proper motions of \textit{Gaia}-CRF3 quasars are roughly an order of magnitude larger than the corresponding powers obtained from simulated proper motions (see the \textit{bottom left panel} of Fig.~\ref{fig:GN_model}). This indicates that the \textit{Gaia}-CRF3 catalog likely contains significant systematic errors, which manifest not only in higher-order harmonics but also substantially affect GW detection. This supports our conclusion from the previous section that both systematic effects and random uncertainties in the proper motions significantly affect the GW detection sensitivity, effectively raising the upper limit on detectable GW amplitudes several times.

The powers $\sqrt{P^s_2}$ and $\sqrt{P^t_2}$ derived from the \textit{Gaia}-CRF3 data, as reported in Table~\ref{tab:vsh_full}, indicate an apparent GW strain amplitude $h_c \sim 4.5\times 10^{-11}$. The statistical significance $Z$ of the dipole and quadrupole powers was computed following \cite{Mignard:2012xm} using the Wilson--Hilferty transformation. Combining the toroidal and spheroidal quadrupole coefficients yields $Z\sim 3.74$, indicating a statistically significant quadrupolar signal, is similar to \cite{Klioner2022} but different from \cite{Darling_2025}. 

Figure~\ref{fig:gaiacrf3} (\textit{bottom left panel}) presents the correlation components $C_{\perp\perp}$ and $C_{\parallel\parallel}$, together with the cross terms ($C_{\parallel\perp}$ and $C_{\perp\parallel}$). Also, the (\textit{bottom right panel}) shows the averaged correlation components and the best-fitting HDC function $\Gamma(\theta)$ and its corresponding $95\%$ confidence interval (\textit{green}), derived after applying a dipole-mode correction based on the proper motions of quasars from the \textit{Gaia}-CRF3 catalog. The fitted HDC amplitude, $\Gamma = 5.92 \pm 4.95 \, \mu as^2 yr^{-2}$, is not statistically significant. This corresponds to a characteristic strain of $h_c \sim 4.1\times10^{-11}$, which is unusually large and, according to the simulations presented in Appendix~\ref{App3_all}, should be detected with a significance exceeding $3\sigma$ in the absence of substantial systematic errors. The dominant higher-order harmonics present in the Gaia DR3 quasar proper-motion field, particularly the $\ell=3$ and $\ell=5$ modes, significantly distort the correlation pattern away from the expected quadrupolar Hellings--Downs form. Consequently, both the shape and amplitude of the recovered correlation function are strongly affected by higher-order systematic and random errors, indicating that the inferred strain amplitude is more likely associated with residual systematics in the Gaia-CRF3 catalog than with a genuine gravitational-wave signal.

Table \ref{tab:vsh_full} lists the estimated VSH coefficients of first and second degree (dipole and quadrupole components). The application of VSH is described in detail, for example, in \cite{Vityazev_2009}. Following this formalism, we decompose the VSH coefficients into their real and imaginary parts, which results in a set of harmonics denoted as $s_{\ell,m,p}$ and $t_{\ell,m,p}$. Here, the index $\ell$ specifies the degree of the harmonic, while $m$ denotes its order. The index $p$ takes two possible values, $p=0$ and $p=1$, corresponding to the imaginary and real parts of the harmonic, respectively. Although the overall amplitudes of these harmonics are broadly consistent with previous studies, we find statistically significant differences with respect to the values reported by \citet{Darling_2025}. These discrepancies are most likely driven by a combination of factors: (i) differences in the adopted quasar selection criteria and sky coverage, and (ii) differences in the treatment of the full astrometric covariance information, in particular the neglect of correlations between the proper motion components in the Quaia catalog \citep{Quaia}. The latter effect is expected to have a non-negligible impact on the estimated amplitudes of VSH harmonics.

\begin{table}
\centering
\small
\caption{VSH coefficients of degree $l=1$ (dipole) and $l=2$ (quadrupole) derived from the 1.5 million \textit{Gaia}-CRF3 quasar sample, in units of $\mu$as yr$^{-1}$.}
\label{tab:vsh_full}
\begin{tabular}{lcc|lcc}
\hline\hline
\multicolumn{3}{c|}{\textbf{B Modes (Toroidal)}} &
\multicolumn{3}{c}{\textbf{E Modes (Spheroidal)}} \\
\hline
Parameter & Value & Unc. & Parameter & Value & Unc. \\
\hline
\multicolumn{6}{c}{\textbf{Dipole ($l=1$)}} \\
\hline
$t_{1,0,1}$ & $-1.219$ & $2.256$ & $s_{1,0,1}$ & $-10.418$ & $1.883$ \\
$t_{1,1,0}$ & $4.202$ & $1.733$ & $s_{1,1,0}$ & $-11.083$ & $1.665$ \\
$t_{1,1,1}$ & $6.763$ & $2.128$ & $s_{1,1,1}$ & $4.161$ & $2.518$ \\
\hline
\multicolumn{3}{c}{$\sqrt{P^t_1}$=$8.05 \pm 3.18$ ($Z=3.05$)} & \multicolumn{3}{c}{$\sqrt{P^s_1}$=$15.77 \pm 3.42$ ($Z=7.47$)}\\
\hline
\multicolumn{6}{c}{\textbf{Quadrupole ($l=2$)}} \\
\hline
$t_{2,0,1}$ & $1.195$ & $2.247$ & $s_{2,0,1}$ & $-0.548$ & $1.916$ \\
$t_{2,1,0}$ & $-1.583$ & $2.350$ & $s_{2,1,0}$ & $7.073$ & $2.097$ \\
$t_{2,1,1}$ & $-2.402$ & $1.653$ & $s_{2,1,1}$ & $-4.049$ & $1.761$ \\
$t_{2,2,0}$ & $-7.091$ & $2.167$ & $s_{2,2,0}$ & $-3.120$ & $2.599$ \\
$t_{2,2,1}$ & $-2.338$ & $1.674$ & $s_{2,2,1}$ & $-2.671$ & $1.691$ \\
\hline
\multicolumn{3}{c}{$\sqrt{P^t_2}$= $8.09 \pm 3.08$ ($Z=2.38$)} & \multicolumn{3}{c}{$\sqrt{P^s_2}$ = $9.14 \pm 3.39$ ($Z=3.08$) } \\
\hline\hline
\end{tabular}
\end{table}

\subsection{Truncated Gaia-CRF3 catalog analysis}
\label{sec:truncatedGaia}

In this subsection, we construct a quasar subsample following the selection criteria presented in \cite{Darling_2025}, i.e., we select all quasars with proper motion amplitudes less than 100~$\mu\mathrm{as}\,\mathrm{yr}^{-1}$. This results in a catalog containing 70,268 quasars. Table~\ref{tab:vsh_truncated} lists the VSH coefficients of the first and second degree (dipole and quadrupole components) derived from this subsample. As shown in Table~\ref{tab:stat_truncated}, the average uncertainty of quasar proper motions is approximately 250~$\mu\mathrm{as}\,\mathrm{yr}^{-1}$.

\begin{table}
\centering
\small
\caption{VSH coefficients of degree $l=1$ (dipole) and $l=2$ (quadrupole) derived from the 70,268 quasar \textit{Gaia}-CRF3 sample, in units of $\mu$as yr$^{-1}$.} 
\label{tab:vsh_truncated}
\begin{tabular}{lcc|lcc}
\hline\hline
\multicolumn{3}{c|}{\textbf{B Modes (Toroidal)}} &
\multicolumn{3}{c}{\textbf{E Modes (Spheroidal)}} \\
\hline
Parameter & Value & Unc. &
Parameter & Value & Unc. \\
\hline
\multicolumn{6}{c}{\textbf{Dipole ($l=1$)}} \\
\hline
$t_{1,0,1}$ & $-2.12$ & $1.90$ &
$s_{1,0,1}$ & $-6.94$ & $1.57$ \\
$t_{1,1,0}$ & $-0.93$ & $1.40$ &
$s_{1,1,0}$ & $1.36$ & $1.33$ \\
$t_{1,1,1}$ & $-3.05$ & $1.74$ &
$s_{1,1,1}$ & $3.36$ & $2.10$ \\
\hline
\multicolumn{3}{c}{$\sqrt{P^t_1}=3.83 \pm 2.93\;(Z=1.57)$} &
\multicolumn{3}{c}{$\sqrt{P^s_1}=7.83 \pm 2.94\;(Z=3.61)$} \\
\hline
\multicolumn{6}{c}{\textbf{Quadrupole ($l=2$)}} \\
\hline
$t_{2,0,1}$ & $2.00$ & $1.83$ &
$s_{2,0,1}$ & $-3.55$ & $1.54$ \\
$t_{2,1,0}$ & $0.24$ & $1.97$ &
$s_{2,1,0}$ & $1.49$ & $1.74$ \\
$t_{2,1,1}$ & $-5.55$ & $1.35$ &
$s_{2,1,1}$ & $-2.08$ & $1.44$ \\
$t_{2,2,0}$ & $-3.69$ & $1.80$ &
$s_{2,2,0}$ & $1.04$ & $2.18$ \\
$t_{2,2,1}$ & $-6.94$ & $1.38$ &
$s_{2,2,1}$ & $-1.53$ & $1.41$ \\
\hline
\multicolumn{3}{c}{$\sqrt{P^t_2}=9.83 \pm 3.77\;(Z=4.87)$} &
\multicolumn{3}{c}{$\sqrt{P^s_2}=4.75 \pm 3.77\;(Z=1.65)$} \\
\hline\hline
\end{tabular}
\end{table}

\begin{table*}
\centering

\caption{Statistical properties of 70,268 selected \textit{Gaia}-CRF3 quasars with proper motion amplitudes below 100 $\mu\mathrm{as}\,\mathrm{yr}^{-1}$.}
\label{tab:stat_truncated}
\begin{tabular}{lccccccc}
\hline
Name & Mean & SD & Minimum & Maximum & Median & Skewness & Kurtosis \\
\hline
$\mu^*_{\alpha}$ ($\mu$as~yr$^{-1}$)         
& 0.0818 & 48.85 & $-99.92$ & 100.00 & 0.234 & $-0.006$ & $-0.95$ \\
$\mu_{\delta}$ ($\mu$as~yr$^{-1}$)          
& $-0.263$ & 48.98 & $-99.89$ & 99.94 & $-0.347$ & 0.002 & $-0.96$ \\
$\sigma_{\mu_{\alpha*}}$ ($\mu$as~yr$^{-1}$)
& 261.82 & 213.79 & 9.72 & 2882.98 & 202.21 & 2.78 & 13.20 \\
$\sigma_{\mu_{\delta}}$ ($\mu$as~yr$^{-1}$) 
& 248.90 & 198.03 & 12.98 & 2695.03 & 195.63 & 2.96 & 15.59 \\
\hline
\end{tabular}
\end{table*}

\begin{figure*}
\includegraphics[width = 90mm]{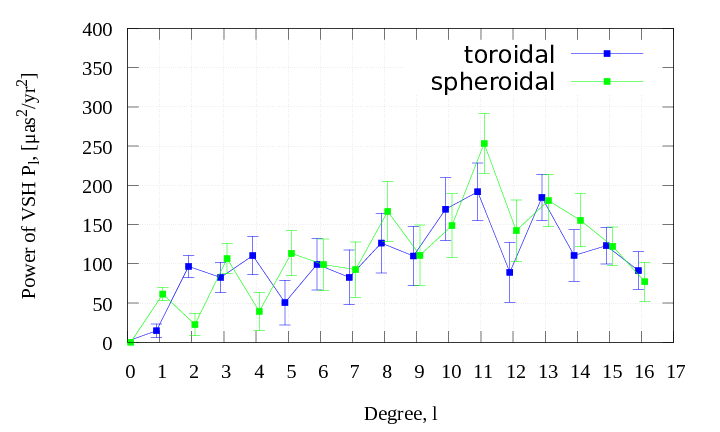}
\includegraphics[width = 90mm]{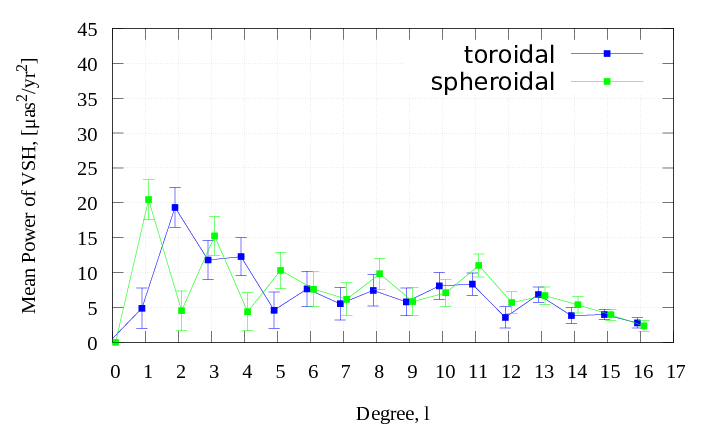}
\includegraphics[width = 90mm]{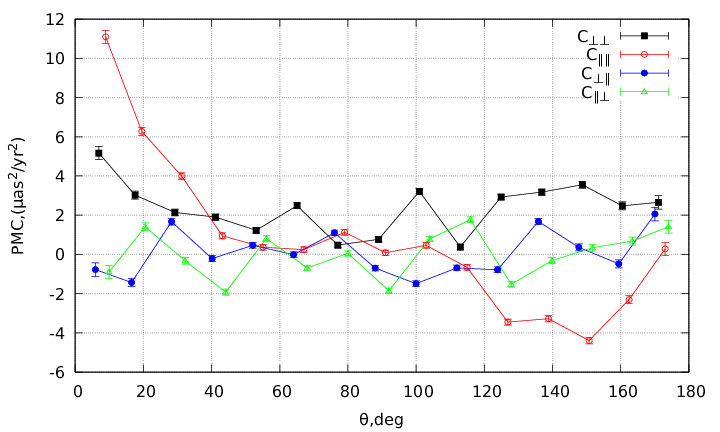}
\includegraphics[width = 90mm]{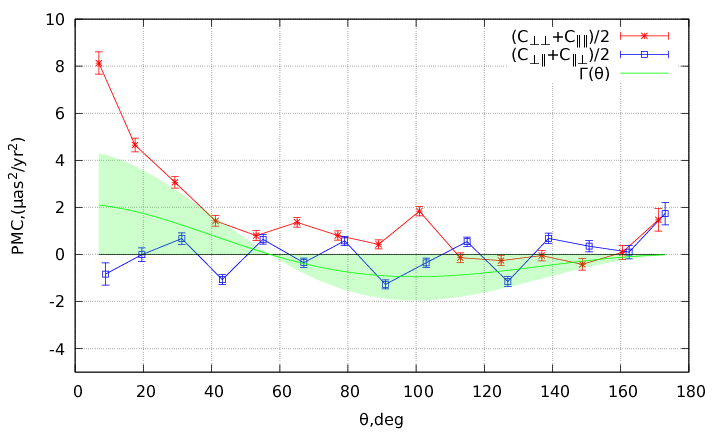}
\caption{ The same that is on Fig. \ref{fig:gaiacrf3} but for the selected 70 268 \textit{Gaia}-CRF3 quasars with proper motion amplitude less than 100 $\mu as \,yr^{-1}$.}
\label{fig:gaiacrf3_cut}
\end{figure*}

As seen in Fig.~\ref{fig:gaiacrf3_cut}, selecting quasars with small proper motion amplitudes preferentially reduces the large scale, low-degree VSH modes ($\ell = 1$--6), which dominate the coherent systematic signal, while leaving higher degree harmonics ($\ell \geq 7$), dominated by small scale noise, largely unaffected. This effect is purely selection driven and results from two factors: (i) the pronounced selection induced non-uniformity in the sky distribution of quasars, primarily dictated by the \textit{Gaia} scanning law, and (ii) the projection of correlated proper motion noise onto low-frequency (low-degree) components. Such a selection suppresses or neglects global systematic patterns in quasar proper motions.

Since the GW signal is encoded predominantly in the lowest-degree VSH, in particular the quadrupolar ($\ell = 2$) modes, selecting small proper motion amplitudes systematically reduces the very components that carry the GW information. Consequently, both VSH- and HDC-based analyzes applied to such a truncated sample yield biased, systematically underestimated values of the GW strain amplitude. Therefore, selection based on proper motion cannot be regarded as statistically unbiased and may lead to misestimation of the GW signal.

As shown in Table~\ref{tab:stat_truncated}, the selected subsample shows a dispersion of proper motions that is significantly smaller than the mean formal uncertainties, together with a negative kurtosis indicative of a truncated distribution. This implies that the applied selection strongly suppresses the tails of the proper motion distribution, resulting in an artificially reduced variance. A kurtosis value close to $-1$ further indicates that the distribution approaches a nearly uniform form within the selected interval rather than a Gaussian one. Such truncation can bias the estimation of GW induced correlations and lead to a systematic underestimation of the recovered GW amplitude, as shown in the analysis of simulated data in Appendices \ref{App2} and \ref{App3}. In Appendix ~\ref{App2:Strict}, we examine how selection criteria applied to simulated quasar proper motions influence both the statistical significance of the detection and the inferred amplitude of the simulated GW signal, as estimated using the VSH and HDC methods. To provide statistically robust validation of the results and conclusions obtained, we additionally performed extensive MC injection and recovery simulations, which are described in detail in Appendix ~\ref{App3}.

\section{Conclusions}
In this work, we investigated the astrometric signatures of GWs in the proper motions of quasars from the \textit{Gaia}-CRF3 catalog. Using realistic simulations based on the actual sky distribution of quasars and their full proper motion covariance matrices, we evaluated the performance of two approaches for astrometric GW searches: the Hellings--Downs correlation analysis and the Vector Spherical Harmonics decomposition.

We developed a covariance-weighted formalism for both methods that accounts for the full uncertainty tensor of quasar proper motions, including the correlations between $\mu_{\alpha *}$ and $\mu_\delta$. Our results show that proper weighting substantially improves the stability and precision of the recovered GW amplitudes. In particular, the uncertainties of both the HDC correlation function and the VSH power spectrum are significantly reduced compared to unweighted analyzes.

The two methods exhibit complementary strengths. The HDC approach provides higher raw statistical sensitivity owing to its quadratic scaling with the number of source pairs, but is more susceptible to anisotropic sky coverage and residual systematics. In contrast, the VSH decomposition provides a mode resolved characterization of the astrometric vector field, facilitating the identification of large scale systematics and anisotropic structures while maintaining robust GW signal recovery. Although HDC and VSH represent different statistical descriptions of the same underlying astrometric signal, they provide mutually consistent constraints on the GW amplitude.

MC injection and recovery simulations performed using the actual \textit{Gaia} DR3 quasar distribution demonstrate that both methods reliably recover injected GW signals for characteristic strains $h_c \gtrsim 10^{-11}$. Below this level, the recovered amplitudes converge toward an approximately constant correlated noise floor, indicating that the practical sensitivity of current astrometric GW searches is limited by residual catalog systematics rather than by statistical noise alone. The simulations further show that the probability of statistically significant detections rapidly decreases below this threshold, while the probability of spurious GW like detections correspondingly increases.

Our analysis also demonstrates that aggressive proper motion selection criteria, such as $|\mu|<0.1~{\rm mas~yr^{-1}}$, introduce significant selection biases. Such cuts systematically suppress the recovered GW amplitudes in both HDC and VSH analyzes by modifying the sky distribution of quasars and enhancing correlations with the \textit{Gaia} scanning law. Consequently, the inferred GW amplitudes may be substantially underestimated despite the apparent reduction of outliers.

The dominant limitation of current astrometric GW searches is therefore not the statistical precision of the estimators themselves, but rather the combined impact of correlated catalog systematics, anisotropic observational noise, uneven sky coverage, and sample-selection effects. This conclusion is further supported by the presence of significant power at multipoles $\ell>2$ in the VSH decomposition of real quasar proper motions, indicating that residual non-GW systematics remain an important component of the observed astrometric signal.

Future \textit{Gaia} data releases will improve the precision of quasar proper motions through longer observational baselines and increased numbers of measurements, thereby increasing the sensitivity of astrometric GW searches. However, further progress will depend on improved mitigation of large-scale astrometric systematics and on statistically optimal likelihood-based analyzes that fully exploit the covariance structure of Gaia-scale datasets. The covariance-weighted methods and MC validation framework presented here provide a basis for future astrometric GW studies and joint \textit{Gaia}--PTA analyzes.

\section{Acknowledgements}
\label{sec:acknowledgements}
This work has made use of data from the European Space Agency (ESA) mission {\it Gaia} (\url{https://www.cosmos.esa.int/gaia}), processed by the {\it Gaia} Data Processing and Analysis Consortium (DPAC, \url{https://www.cosmos.esa.int/web/gaia/dpac/consortium}). Funding for the DPAC has been provided by national institutions, in particular the institutions participating in the {\it Gaia} Multilateral Agreement. 
L.F., M.C., U.A.  acknowledge support from the PRIN 2022 grant 20227MYL2X “General Relativistic Astrometry and Pulsar Experiment (GRAPE) financed by the European Union - Next Generation EU, Mission 4 Component 1 CUP C53D23000890006.  
The authors thank the anonymous reviewer for helpful corrections, comments, and suggestions.

\section{Data availability}
\addcontentsline{toc}{section}{Data availability}

The catalog data used are available in a standardized format for readers via the CDS (https://cds.u-strasbg.fr). 
The software code used in this paper can be made available on personal request by e-mail: \href{mailto:akhmetovvs@gmail.com} {akhmetovvs@gmail.com}.

%
\bibliographystyle{aa} 
\bibliography{GW_QSO.bib} 
%

\begin{appendix}
\section{Simulation of a Stochastic Gravitational-Wave Background}
\label{App1}
To assess the astrometric response of a quasar sample to a SGWB, we performed dedicated simulations in which the background was modeled as a superposition of a large number of independent GWs generated by a population of supermassive black hole binaries (SMBHBs), as these are the most promising SGWB candidates. This population was created from a simulated catalog compatible with PTA data\cite{valtolina}.

Roughly 120 \,000 binaries were quasi-uniformly distributed on the sky, and from each source a contribution to the background was calculated from the physical parameters of the catalog (i.e. chirp mass, frequency, physical distance, and inclination) plus a random phase. The total GW signal was then constructed as the linear superposition of the plane waves produced by all emitters. As a result, we obtained a time-dependent, spatially varying GW-induced deformation field corresponding to an effectively isotropic source distribution.

\begin{figure*}
\includegraphics[width = 90mm]{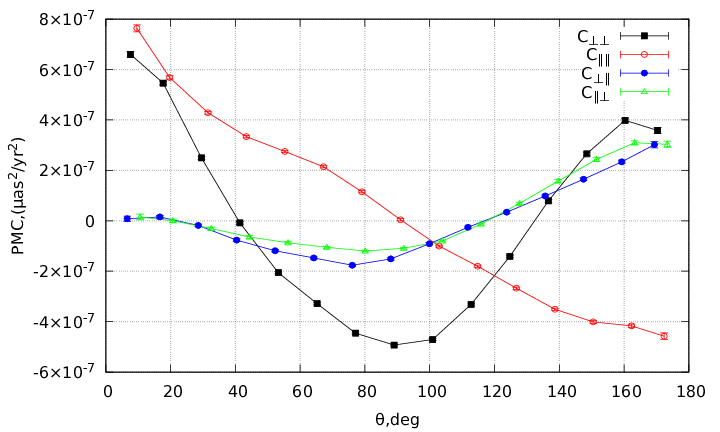}
\includegraphics[width = 90mm]{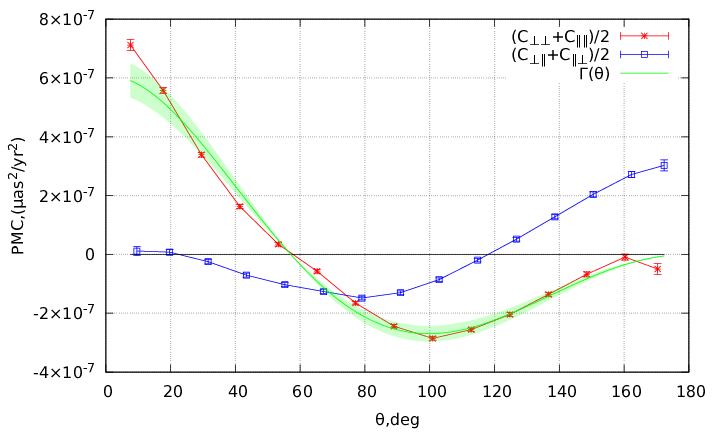}
\includegraphics[width = 90mm]{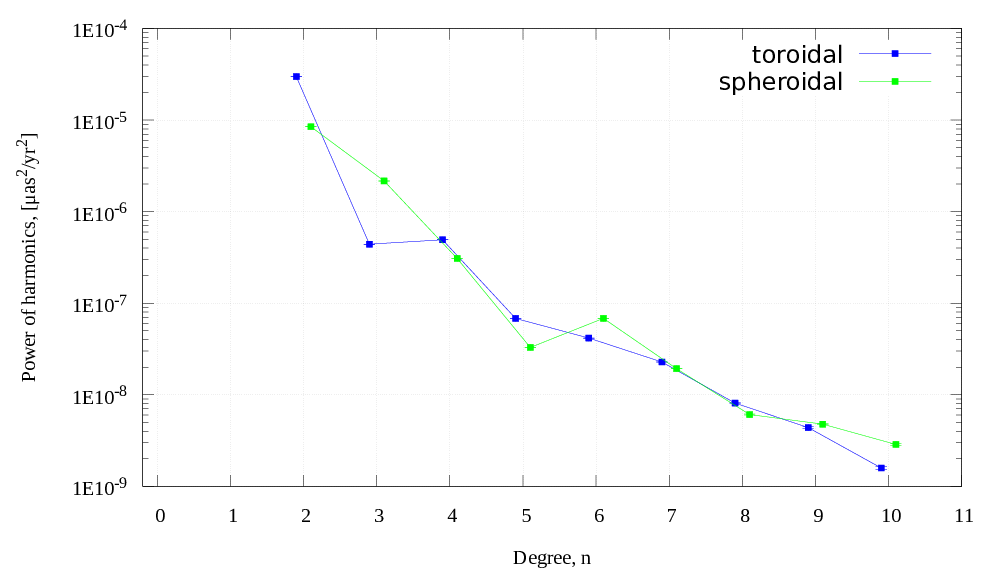}
\includegraphics[width = 90mm]{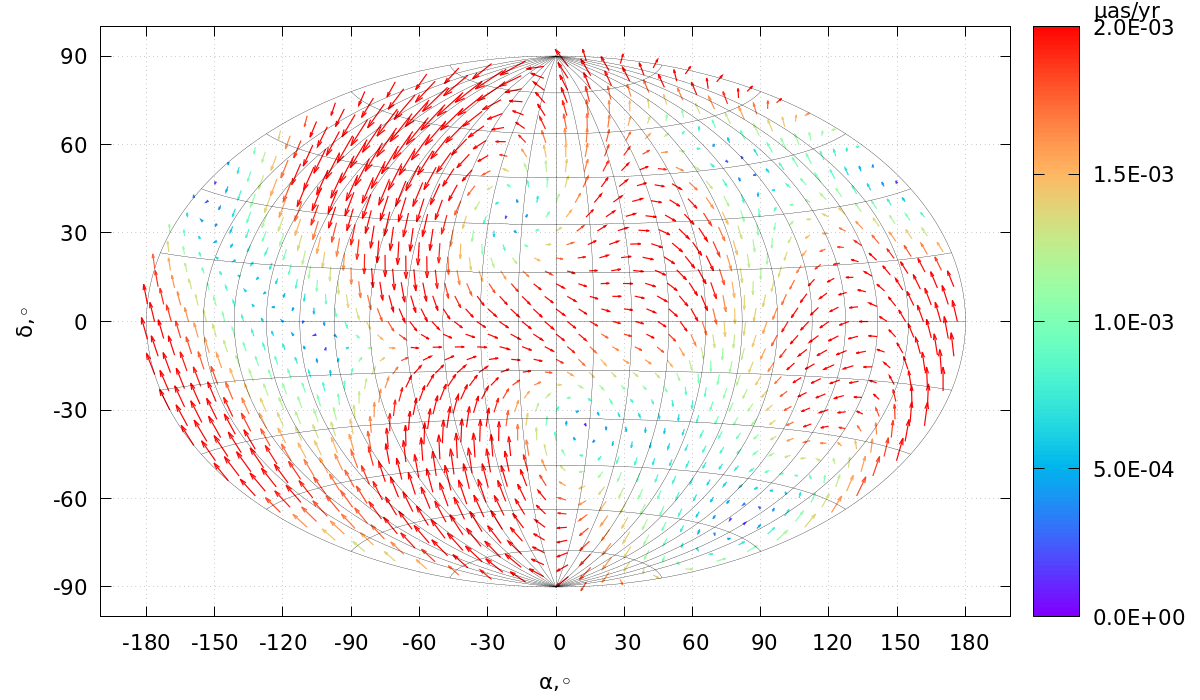}
\caption{The \textit{top left row} presents the proper motion correlation components derived using the HDC method for all four PMC components, computed from the simulated SGWB.
The \textit{top right panel} shows the averaged correlation components $C_{\perp\perp}$ and $C_{\parallel\parallel}$ (\textit{red}), together with the cross terms $C_{\parallel\perp}$ and $C_{\perp\parallel}$ (\textit{blue}). Also shown are the best-fitting HDC function $\Gamma(\theta)$ and its corresponding $95\%$ confidence interval (\textit{green}), derived from these correlation curves.
The \textit{bottom left panel} displays the power spectrum of the toroidal (\textit{blue points}) and spheroidal (\textit{green points}) VSH obtained from the simulated SGWB.
The expected proper motion field resulting from the simulated SGWB is shown in the \textit{bottom right panel}.
The simulations were performed without including observational uncertainties in quasar proper motions in order to estimate the intrinsic amplitude of the geometric distortions induced by the SGWB.}
\label{fig:SGWB}
\end{figure*}

The induced apparent proper motions of quasars were computed from the resulting metric perturbations. The expected proper motion field obtained from this superposition is shown in Fig. \ref{fig:SGWB} (\textit{lower right panel}). The \textit{upper panel} of Fig. \ref{fig:SGWB}  presents the proper motion correlation components derived using the HDC method.
As seen in Fig. \ref{fig:SGWB}, even for a uniform distribution of GW sources across the sky, the correlations do not follow the theoretical slope. The cross-correlation terms ($C_{\parallel\perp}$ and $C_{\perp\parallel}$) do not vanish at all angular separations due to the anisotropic spatial distribution of quasars in optical catalogs. In principle, for future detectors, such an effect might become an insurmountable obstacle for the detection of GWB anisotropies with astrometry: implementation of more quantitative and robust analysis will be developed in future works. 

Moreover, also the parallel and perpendicular correlation components ($C_{\parallel\parallel}$ and $C_{\perp\perp}$) deviates from the expected HDC with two different slopes (although they should have identical behavior). However, taking the average values mitigates these effects caused by quasar spatial distribution and cosmic variance: as shown in the \textit{upper right panel} in figure \ref{fig:SGWB}, the fit can reconstruct the correct characteristic strain from the $(C_{\parallel\parallel} +C_{\perp\perp})/2$ correlation term, showing that the large-scale geometric signature of the background remains dominated by the quadrupolar correlation structure characteristic of GWs.

The \textit{lower left panel} of Fig. \ref{fig:SGWB} shows the power spectrum of the toroidal (\textit{blue}) and spheroidal (\textit{green}) points. The distribution of power across modes confirms that the resulting signal retains the characteristic low-degree structure expected from the plane GW.

We emphasize the fact that such catalog was designed for PTA periodic signal detection, therefore some binaries will have an orbiting frequency higher than current Gaia DR3 frequency upper limit. However, here we are not interested in a fully realistic scenario, but rather we want to estimate the geometric structure of astrometric distortion and the effects that the anisotropic distribution of quasars might have on SGWB induced observables. This approach provides a practical upper-limit estimate of the level of correlated proper motion patterns that could be present in real astrometric data.

\section{Checking of selection criteria and weighting scheme}
\label{App2}

In this appendix, we investigate how different selection criteria and weighting schemes applied to quasar proper motions affect the recovered GW signal when using both the HDC and VSH methods.

To this end, we simulate a plane GW from a single source with the parameters specified in Sect.~\ref{sec:simulation}. Gaussian noise is then added to the simulated proper motions. Noise is modeled using the covariance matrices of the \textit{Gaia}-CRF3 proper motions, taking into account both the standard uncertainties of the proper motion components and their correlations (Sect.~\ref{subsec:Noise}).

To obtain a more reliable estimate of the amplitude of the simulated GW signal and to approximate the expected improvement in the forthcoming \textit{Gaia} DR4 release, the noise level is reduced by a factor of three, i.e.,
$ \sigma_{\mu_\alpha} \rightarrow 0.3\,\sigma_{\mu_\alpha}$ and $\sigma_{\mu_\delta} \rightarrow 0.3\,\sigma_{\mu_\delta}$. The statistical properties of the simulated proper motions and their corresponding uncertainties for 1.5 million quasars are listed in Table~\ref{tab:B0}. As seen in the table, the standard deviations of the simulated proper motions are close to the mean uncertainties reported in the \textit{Gaia}-CRF3 catalog, which have been reduced by a factor of three. Moreover, the skewness and kurtosis of the simulated proper motion distribution closely the corresponding values derived from the real data, as listed in Table~\ref{tab:quasar_stats}. This indicates that the simulated data set realistically reproduces the statistical characteristics of the observed quasar sample, making it suitable for testing analysis methods such as HDC and VSH. 

\subsection{Comparison of weighted and unweighted approaches}
\label{App2:weighting}

In this subsection, we investigate how the application of statistical weights affects the estimation of the GW amplitude derived from simulated quasar proper motions. In particular, we compare two approaches: (i) an unweighted analysis, in which all quasars contribute equally to the statistics in both the HDC and VSH methods, and (ii) a weighted analysis, in which each measurement is weighted according to its formal uncertainty.

The weighted scheme accounts for the heterogeneous astrometric precision of the sources in the \textit{Gaia}-CRF3 catalog. In practice, quasars with smaller proper motion uncertainties receive larger statistical weights, while objects with larger uncertainties contribute less to the final estimator.

In the HDC framework, the proposed weighting scheme allows different weights to be assigned to the proper motion components projected along the directions parallel and perpendicular to the great circle connecting each pair of sources. As a result, the same quasar may have different effective weights depending on the specific source pair and on the projected component of the proper motion considered in the correlation analysis (formulas \ref{eq:sig}, \ref{eq:varianc} and \ref{eq:C_par_per_w}).

\begin{figure*}
\includegraphics[width = 90mm]{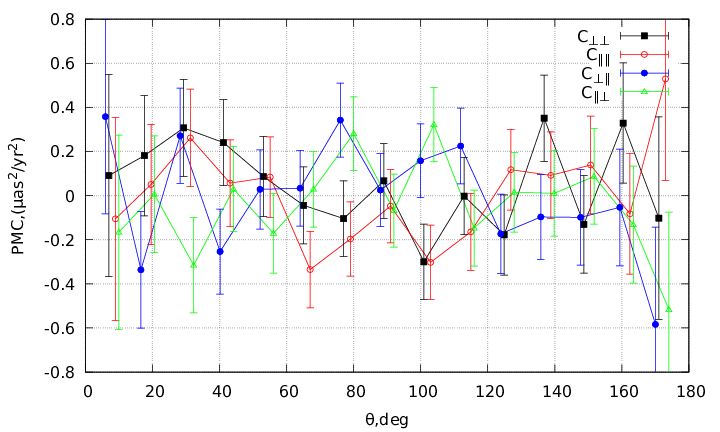}
\includegraphics[width = 90mm]{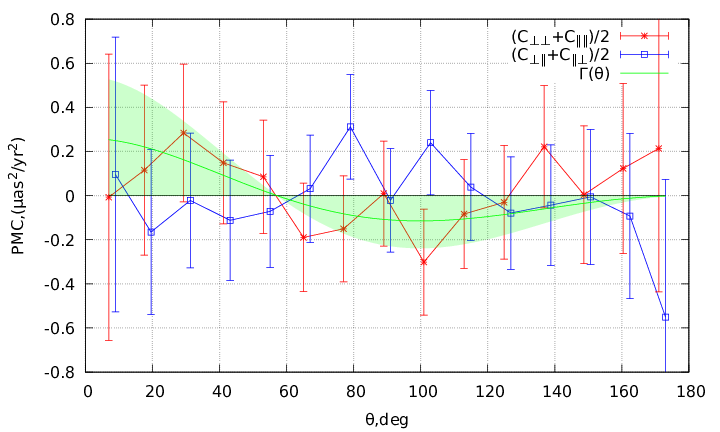}
\includegraphics[width = 90mm]{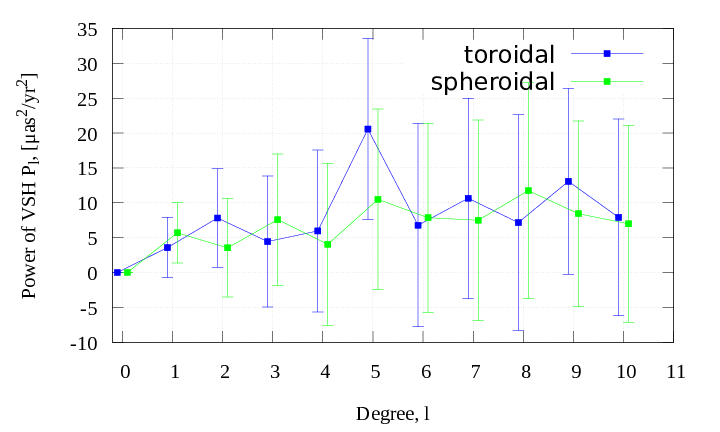}
\includegraphics[width = 90mm]{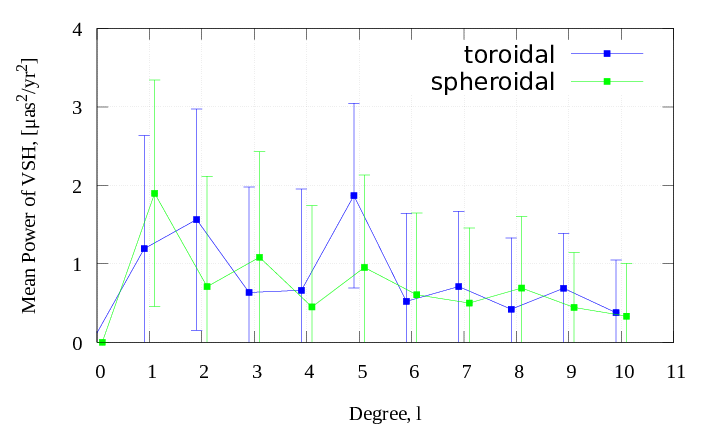}
\caption{ Unweighted analysis. The \textit{top row} shows HDC for all four PMC components from simulated proper motions with the \textit{Gaia}-CRF3 error model (\textit{top left}), and the averaged components $C_{\perp\perp}$, $C_{\parallel\parallel}$ (\textit{red}) with cross terms $C_{\parallel\perp}$, $C_{\perp\parallel}$ (\textit{blue}) and the fitted HD function $\Gamma(\theta)$ with 95 \% confidence interval (\textit{green}, \textit{top right}).
The \textit{bottom row} shows the power spectra of VSH coefficients (\textit{left}) and mean powers (\textit{right}) for toroidal (\textit{blue}) and spheroidal (\textit{green}); coefficients are offset by $\Delta \ell = 0.1$ for clarity.
Simulations assume a plane GW with equal “$+$” and “$\times$” polarizations, $h_c = 10^{-11}$, propagating toward $\alpha = 45^\circ, \delta = 45^\circ$, with the random proper motion component reduced by a factor of three ($0.3\sigma_{\mu^*_\alpha}, 0.3\sigma_{\mu_\delta}$), for a sample of 1\,505\,427 quasars.} 
\label{fig:B0}
\end{figure*}

To quantify the impact of weighting, we apply both approaches to the same set of simulated proper motions described in Sect.~\ref{sec:simulation}. In the \textit{top panel} of Figure~\ref{fig:B0}, the HDC curves are shown, derived from simulated GWs with added noise, using the actual \textit{Gaia}-CRF3 quasar positions together with their proper motion uncertainties and correlations.
When unweighted statistics of quasar proper motions are used, neither the HDC nor the VSH method provides a reliable recovery of the model GW amplitude. The uncertainty of the inferred function $\Gamma$, derived from the proper motion correlation components, exceeds its estimated value, which is nearly half the injected model amplitude (\textit{top right panel}).
From the \textit{bottom panel} of Figure~\ref{fig:B0}, it is also evident that the recovered powers of all harmonics do not exceed their corresponding uncertainties and remain in the range of $5$ to $15\,\mu\mathrm{as}^2\,\mathrm{yr}^{-2}$. Moreover, high-order harmonics dominate the power spectrum, indicating a highly non-uniform, high-frequency distribution of proper motion errors across the sky, which is associated with the \textit{Gaia} scanning law.

\begin{figure*}
\includegraphics[width = 90mm]{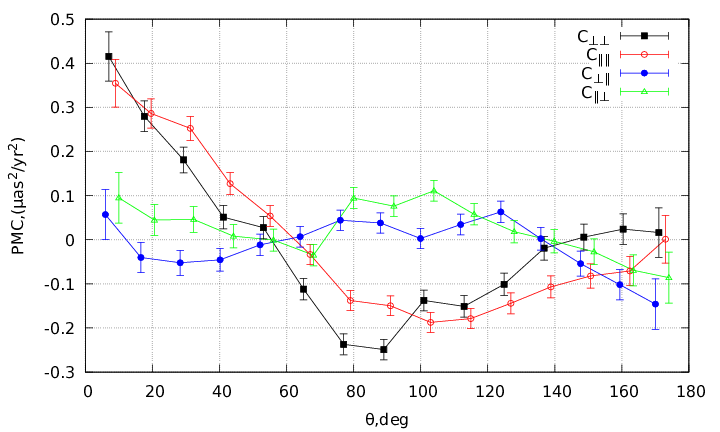}
\includegraphics[width = 90mm]{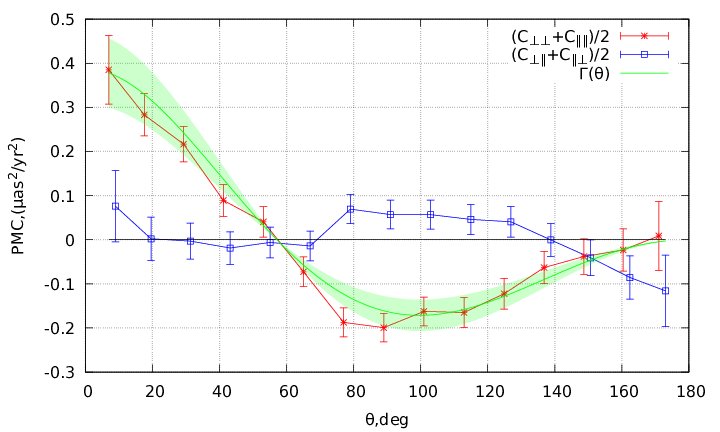}
\includegraphics[width = 90mm]{figures/power_harmonics_s0.0_r3.0_w2_gen_Noise_GW_model_gaiacrf3_k0.3.png}
\includegraphics[width = 90mm]{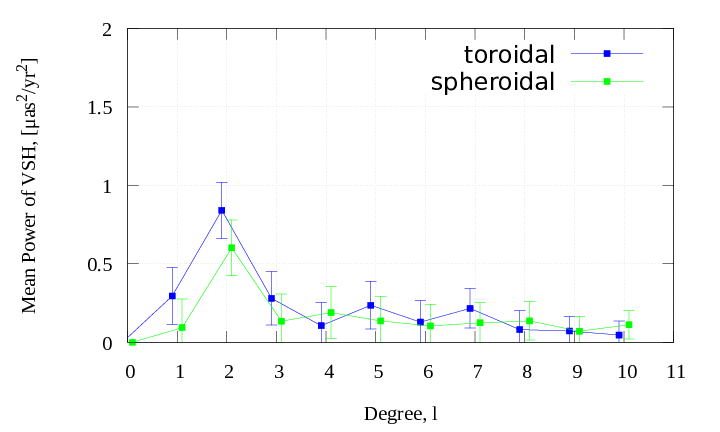}
\caption{Same as Fig. \ref{fig:B0}, but applying statistical weights to the same dataset of quasar proper motions.} 
\label{fig:B0w}
\end{figure*}

The \textit{top panels} of Figure~\ref{fig:B0w} show the HDC components, while the \textit{bottom ones} show the power spectrum of the VSH harmonics obtained by applying statistical weights to the exact same simulated sample of quasar proper motions as in Figure~\ref{fig:B0}. It is evident that using the correct weighting scheme, which fully accounts for the covariance matrix of quasar proper motions, allows a reliable recovery of the model GW amplitude. The uncertainty of the inferred $\Gamma(\theta)$ function is less than 10\% of its value and the value is consistent with the injected model amplitude. 
Similarly, among the recovered VSH powers, the quadrupolar component $P_2$ clearly dominates over the higher-order harmonics and closely matches the expected theoretical value. The uncertainties in both the correlation components and the VSH powers are nearly an order of magnitude smaller than in the unweighted case. 
This demonstrates that the correct weighting of quasar proper motions is essential for an accurate estimation of the GW amplitudes and for maximizing the sensitivity of both the HDC and VSH methods. In addition, correct weighting mitigates spurious contributions arising from non-uniform sky coverage, reduces the impact of high-frequency noise, and ensures that large-scale systematic patterns in the proper motion field are properly captured.

\begin{table*}
\centering
\caption{Statistical properties of the simulated proper motions for 1 505 427 quasars, with the noise level reduced by a factor of three relative to the \textit{Gaia}-CRF3 catalog.}
\label{tab:B0}
\begin{tabular}{lccccccc}
\hline
Name & Mean & SD & Minimum & Maximum & Median & Skewness & Kurtosis \\
\hline
$\mu^*_{\alpha}$ ($\mu$as~yr$^{-1}$)  & 0.0633 & 250 & -3301 & 3607 & 0.0460 & -0.00524 & 7.46 \\
$\mu_{\delta}$ ($\mu$as~yr$^{-1}$)  & 0.104  & 233 & -3044 & 3136 & 0.135  & -0.0230 & 7.97 \\
$\sigma_{\mu^*_{\alpha}}$ ($\mu$as~yr$^{-1}$) & 201 & 148 & 2.92 & 1010 & 158 & 1.583 & 2.872 \\
$\sigma_{\mu_{\delta}}$ ($\mu$as~yr$^{-1}$) & 187 & 138 & 3.89 & 1016 & 147 & 1.671 & 3.396 \\
\hline
\end{tabular}
\end{table*}

\subsection{Strict proper motion selection}
\label{App2:Strict}
We apply a stringent selection criterion, analogous to that adopted by \citep{Darling_2025}. Since the simulated noise level in our model is reduced by a factor of three relative to \textit{Gaia}-CRF3, we impose a proportional constraint on the proper motion amplitudes in order to preserve the ratio $|\mu|/\sigma_\mu$. 

Specifically, we apply selection criteria to the previously simulated dataset, retaining only quasars with total proper motion amplitudes $\leq 33~\mu$as\,yr$^{-1}$. This yields a sample of 84,607 objects, comparable in size to both the subset analyzed in Sect.~\ref{sec:truncatedGaia} and that used by \cite{Darling_2025}. This allows us to assess, using simulated data, how such a strict selection affects the recovered GW signal.

\begin{figure*}
\includegraphics[width = 90mm]{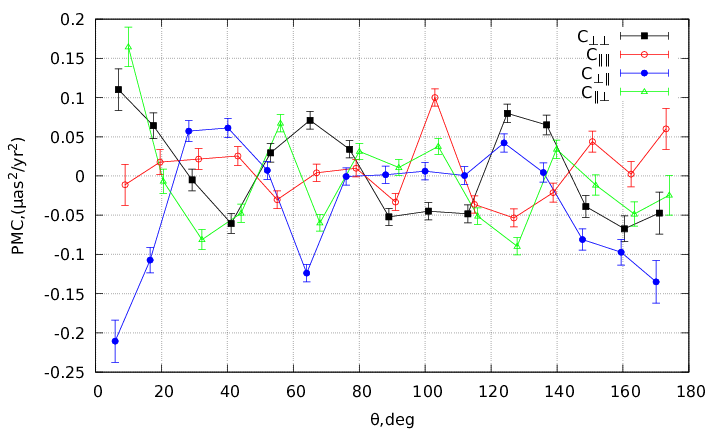}
\includegraphics[width = 90mm]{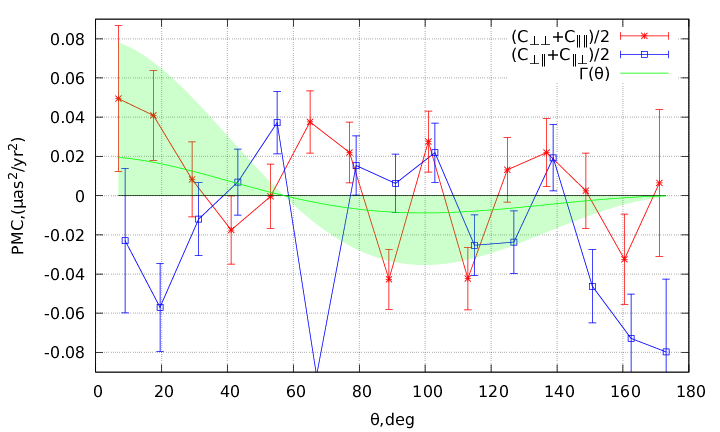}
\includegraphics[width = 90mm]{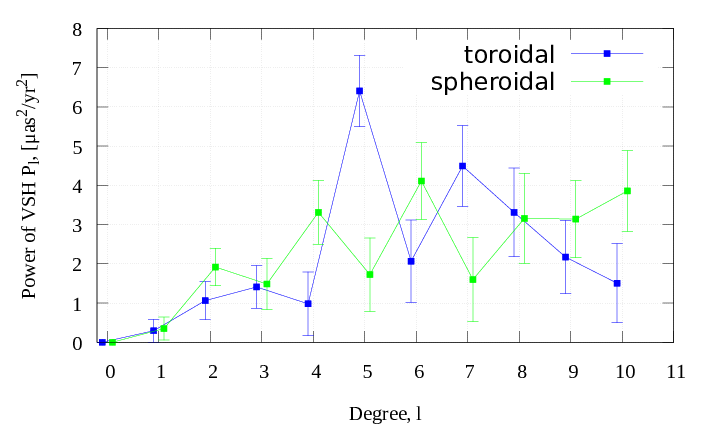}
\includegraphics[width = 90mm]{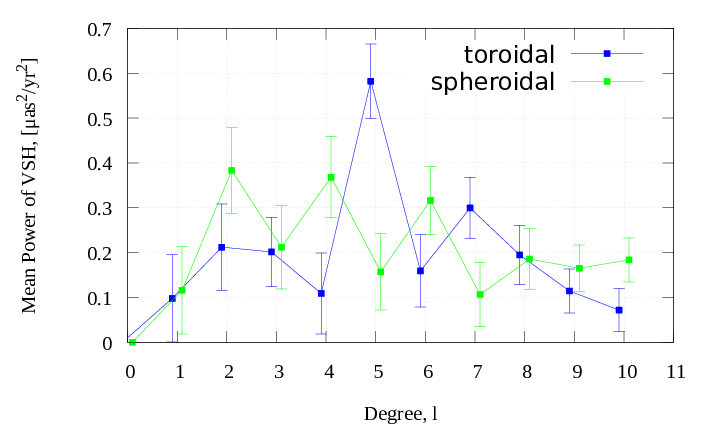}
\caption{Same as Fig. \ref{fig:B0}, using statistical weights for the selected 84\,607 quasars with simulated proper motion amplitudes less than 33~$\mu\mathrm{as}\,\mathrm{yr}^{-1}$.
}
\label{fig:B2}
\end{figure*}

The statistical properties of this subsample are listed in Table \ref{tab:B2}. The upper panel of Fig.\ref{fig:B2} shows the recovered correlation components and the corresponding HD curve. The recovered HDC amplitude is suppressed by a factor of 20, meaning that the inferred GW strain amplitude is underestimated by a factor of nearly 4.5 (specifically, $\sqrt{20}$).

In contrast, the lower panel of Fig.\ref{fig:B2} presents the VSH power spectrum. The quadrupolar ($\ell = 2$) harmonic remains suppressed by approximately a factor of two relative to the expected model value, corresponding to an underestimation of the GW amplitude by roughly $\sqrt{2}$. However, the power of higher-order harmonics, particularly $\ell = 5$ and $\ell = 7$, increases by a factor of 2--3. This indicates a pronounced non-uniformity in the sky distribution and in the error properties of the selected sample, leading to the appearance of spurious high-order modes.

These results demonstrate that strong selection cuts on proper motion amplitude introduce significant non-uniformity in the sky distribution of quasars and systematically suppress the low-degree harmonics that encode the GW signal. 

The VSH method appears more robust to such selection effects than the HDC approach, as it better isolates the quadrupolar component from higher-order contamination. Nevertheless, severe restrictions of the type $ |\mu| / \sigma_\mu< 3$ can still lead to a systematic underestimation of the GW strain amplitude.

We therefore conclude that proper motion based selection must be applied with caution in GW searches using quasar astrometry, as overly strict cuts may bias the inferred GW amplitude. The Monte Carlo simulations presented in Appendix \ref{App3} confirm this effect, showing that severe truncation of the proper-motion distribution leads to a systematic underestimation of the injected GW strain, even in the absence of additional systematics.

\begin{table*}
\centering
\caption{Statistical properties of the selected sample of 84\,607 quasars, with the noise level reduced by a factor of three relative to the \textit{Gaia}-CRF3 catalog. The simulated proper motions are less than 33~$\mu\mathrm{as}\,\mathrm{yr}^{-1}$.}
\label{tab:B2}
\begin{tabular}{lccccccc}
\hline
Name & Mean & SD & Minimum & Maximum & Median & Skewness & Kurtosis \\
\hline
$\mu^*_{\alpha}$ ($\mu$as~yr$^{-1}$)  & 0.086 & 15.4 & -31.6 & 31.6 & 0.083 & 1.58e-4 & -0.937 \\
$\mu_{\delta}$ ($\mu$as~yr$^{-1}$)  & 0.029 & 15.4 & -31.6 & 31.6 & 0.0395 & -0.00367 & -0.940 \\
$\sigma_{\mu^*_{\alpha}}$ ($\mu$as~yr$^{-1}$)  & 79.2 & 65.1 & 2.92 & 921 & 60.9 & 2.94 & 15.3 \\
$\sigma_{\mu_{\delta}}$ ($\mu$as~yr$^{-1}$)  & 75.2 & 59.8 & 3.88 & 973 & 59.1 & 3.02 & 16.3 \\
\hline
\end{tabular}
\end{table*}

\section{Monte Carlo injection and recovery simulations}
\label{App3}

To further assess the robustness and sensitivity of the HDC and VSH analyzes under realistic \textit{Gaia} observational conditions, we performed a set of Monte Carlo injection and recovery simulations using the actual sky distribution and astrometric uncertainties of approximately $1.5$ million quasars from the \textit{Gaia} DR3 catalog.

For each realization, synthetic proper motions were generated by injecting plane GW signals with characteristic strains in the range $10^{-12} \leq h_c \leq 10^{-10}$, using randomly generated GW phases and sky locations. Correlated Gaussian noise was added according to the full covariance matrices of the \textit {Gaia}-CRF3 proper motions, including both the formal uncertainties and the correlations between the proper motion components. In total, 1000 independent Monte Carlo realizations were generated over the selected range of injected GW amplitudes.

\subsection{Monte Carlo simulations based on all \textit {Gaia}-CRF3 data}
\label{App3_all}

\begin{figure*}
\includegraphics[width = 93 mm]{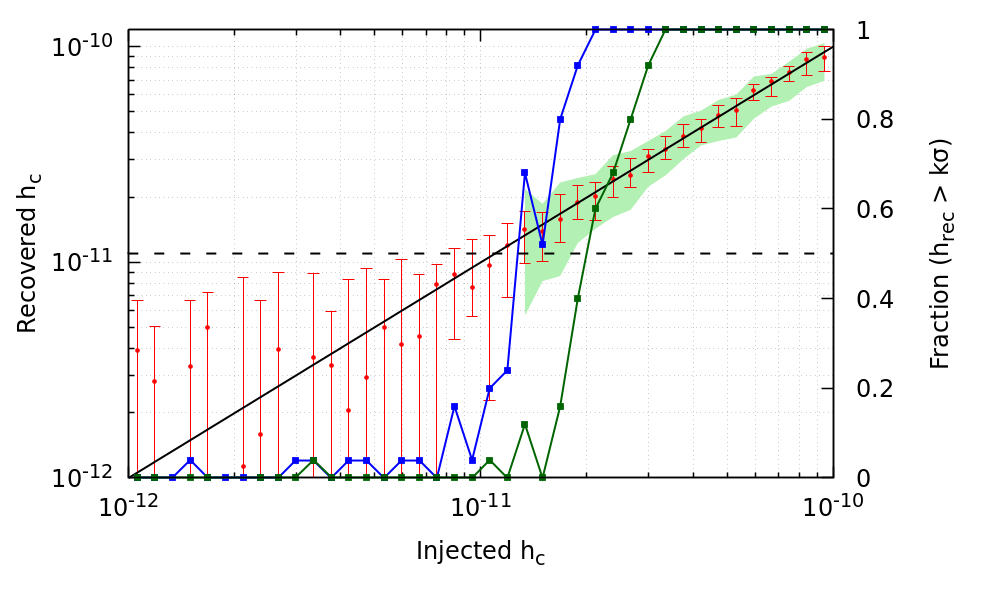}
\includegraphics[width = 93 mm]{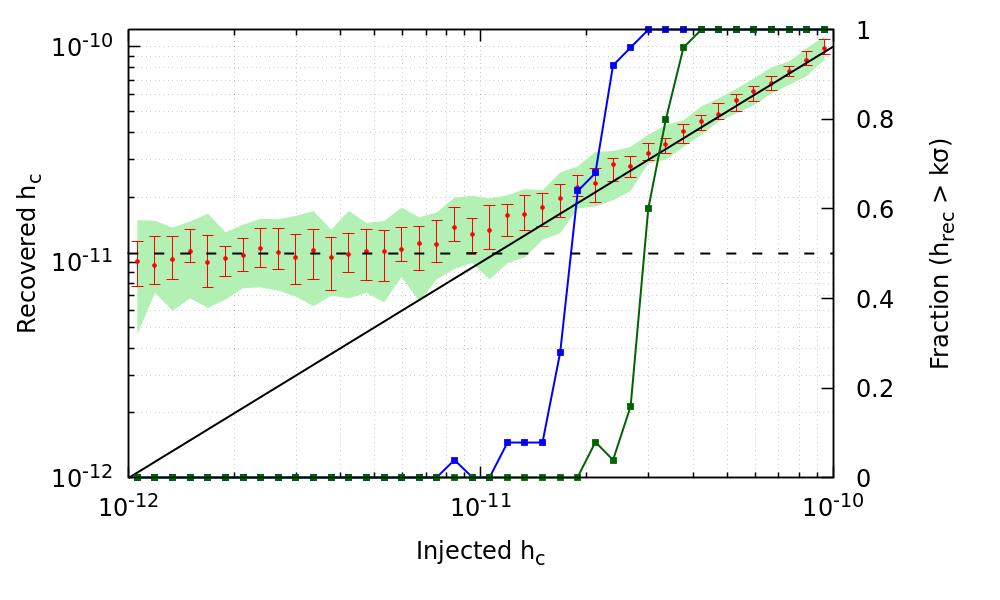}
\caption{Monte Carlo injection and recovery simulations of the GW strain amplitude using the HDC (\textit{left}) and VSH (\textit{right}) methods. Each point shows the median recovered amplitude. Error bars indicate the central 15th--85th percentile interval of the recovered distributions, while the \textit{green shaded region} represents the 95\% confidence interval (2.5th--97.5th percentiles). The \textit{black} diagonal line denotes the ideal relation $h_{rec}=h_{inj}$. The \textit{blue} and \textit{dark-green} curves (right axis) show the fractions of realizations satisfying $h_{rec}>2 \sigma$ and $h_{rec}>3 \sigma$, respectively.}
\label{fig:C1}
\end{figure*}

Figure \ref{fig:C1} presents the results of the Monte Carlo injection and recovery simulations for the plane GW strain amplitude using the HDC (\textit{left}) and VSH (\textit{right}) methods applied to the full \textit{Gaia} DR3 quasar sample containing approximately 1.5 million sources. The horizontal axis shows the injected characteristic strain amplitude $h_c$, while the vertical axis corresponds to the recovered value obtained from the HDC fit and the VSH power spectrum analysis. Each point represents the median recovered amplitude derived from 40 independent Monte Carlo realizations performed with random GW phases and sky locations. The red error bars indicate the central 15th--85th percentile interval of the recovered distributions, while the green shaded region corresponds to the $2.5-97.5$ percentile range. The black diagonal line marks the ideal one to one relationship between injected and recovered amplitudes. 

Simulations demonstrate that for injected amplitudes $h_c \gtrsim 10^{-11}$, the recovered values closely follow the expected linear relationship, indicating that both HDC and VSH methods can reliably reconstruct GW signals above this level under realistic \textit{Gaia} DR3 observational conditions.

Additional information is provided by the blue and green curves shown on the right-hand axis of Fig.~\ref{fig:C1}, which represent the fractions of MC realizations satisfying $h_{\rm rec}>2\sigma_{h_{\rm rec}}$ and $h_{\rm rec}>3\sigma_{h_{\rm rec}}$, respectively. These quantities provide empirical estimates of the probability of obtaining statistically significant GW detections at confidence levels $2\sigma$ and $3\sigma$. For a given injected amplitude $h_c$, the detection fractions are defined as
\begin{equation}
f_{n\sigma}(h_c)=
\frac{N\!\left(h_{\rm rec}>n\,\sigma_{h_{\rm rec}}\right)}
     {N_{\rm MC}},
\end{equation}
where $N_{\rm MC}$ is the total number of MC realizations and
$N(h_{\rm rec}>n\,\sigma_{h_{\rm rec}})$ is the number of realizations for which the recovered GW amplitude exceeds $n$ times its estimated uncertainty.

The dashed horizontal line corresponds to $f_{n\sigma}=0.5$, indicating the injected GW amplitude at which more than $50\%$ of the realizations yield a statistically significant detection. This criterion provides a convenient empirical estimate of the practical detection threshold.

For injected amplitudes $h_c \gtrsim 10^{-11}$, both fractions rapidly increase toward unity, indicating that the HDC and VSH methods recover the GW signal with high statistical significance in the majority of realizations. In contrast, for weaker injected signals the detection fractions decrease substantially, reflecting the growing influence of correlated astrometric noise and residual catalog systematics. In the limit of very small injected amplitudes, the non-zero values of these fractions provide an estimate of the probability of obtaining spurious GW-like detections produced by correlated noise fluctuations alone. The transition region around $h_c \simeq 1.15\times10^{-11}$ therefore marks the practical sensitivity threshold of the current \textit{Gaia} DR3 data, where the detection efficiency rapidly declines and the probability of false detections becomes non-negligible.

However, below $h_c \sim 10^{-11}$, the recovery rapidly departs from the ideal relation and approaches an approximately constant floor near $h_c \sim (0.5-1.0)\times10^{-11}$, as shown in Fig.~\ref{fig:C1}. This behavior indicates the presence of an effective correlated noise limit dominated by residual astrometric systematics, anisotropic proper motion uncertainties, scanning law, and spatially correlated errors in the \textit{Gaia} reference frame. In this weak signal regime, the injected GW signal becomes indistinguishable from the correlated noise background, leading to saturation of the recovered amplitudes and a substantial increase in relative uncertainties. Consequently, fluctuations produced by correlated astrometric noise alone may mimic a weak GW signal, increasing the probability of false detections near the sensitivity threshold. Figure~\ref{fig:C1} therefore suggests that the practical sensitivity floor for GW detection with the current \textit{Gaia} DR3 proper-motion uncertainties and covariance structure is close to $h_c \sim 10^{-11}$, below which reliable discrimination between genuine GW signals and correlated astrometric noise becomes increasingly difficult.

\subsection{Monte Carlo simulations with strict proper motion selection}
\label{App3_cut}

\begin{figure*}
\includegraphics[width = 93 mm]{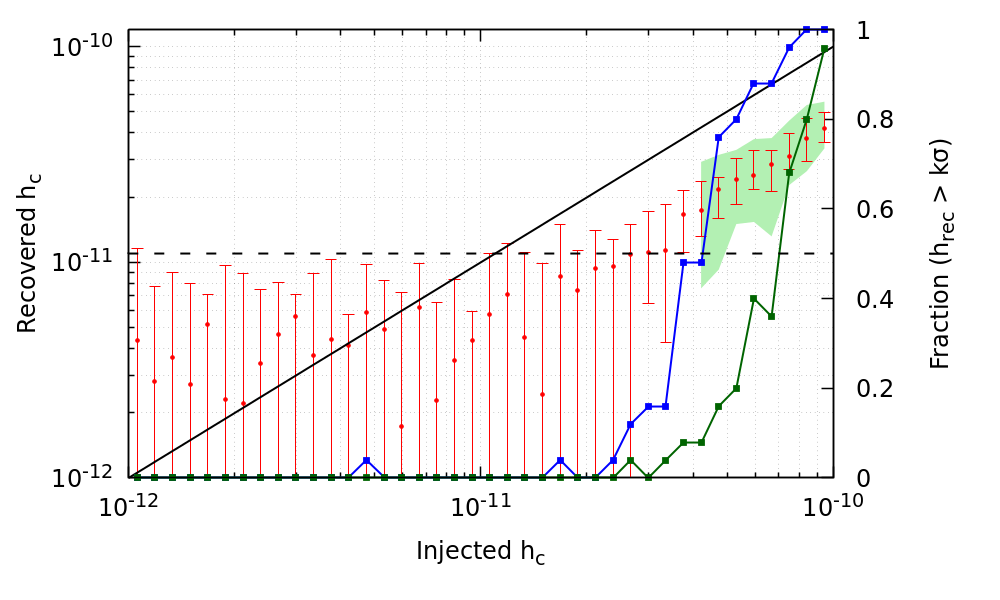}
\includegraphics[width = 93 mm]{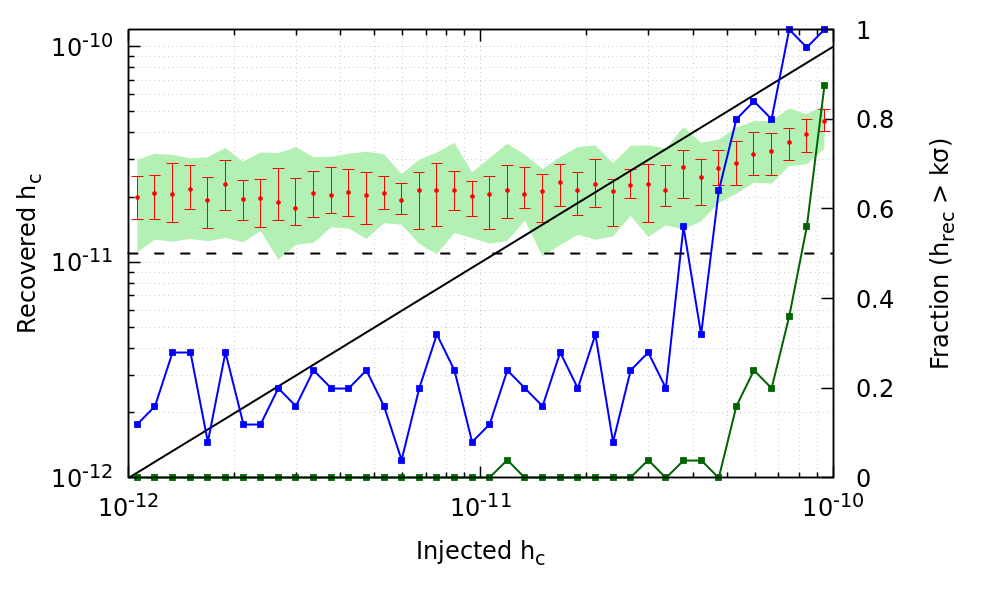}
\caption{Same as Fig.~\ref{fig:C1}, but for the subsample with simulated proper motion amplitudes restricted to $|\boldsymbol{\mu}| < 100~\mu$as,yr$^{-1}$ using the HDC (\textit{left}) VSH (\textit{right}) methods.}
\label{fig:C2}
\end{figure*}

Figure~\ref{fig:C2} presents the Monte Carlo injection and recovery simulations performed after applying the additional proper motion selection criterion $|\mu|<0.1\ {\rm mas\ yr^{-1}}$, analogous to the sample selection adopted by \citet{Darling_2025}. As in Fig.~\ref{fig:C1}, the left and right panels show the results obtained with the HDC and VSH methods, respectively.

A comparison with the full sample results reveals a pronounced systematic suppression of the recovered GW amplitudes. Unlike the nearly unbiased recovery obtained for the complete \textit{Gaia} DR3 quasar sample, both the HDC and VSH analyzes now underestimate the injected signal over the entire amplitude range. Even for relatively strong injected signals, $h_c\sim10^{-10}$, the recovered amplitudes remain significantly below the ideal one-to-one relationship.

The bias becomes particularly severe in the weak signal regime. For injected amplitudes below approximately $h_c\sim4\times10^{-11}$, the recovered amplitudes show only a weak dependence on the injected signal strength and remain clustered around an approximately constant level of $(1-2)\times10^{-11}$. This behavior indicates that the reconstruction becomes dominated by correlated astrometric noise and residual systematic effects, preventing the reliable recovery of weaker GW signals.

The detection efficiency curves shown on the right-hand axes provide additional insight into the statistical significance of the recovered signals. For both HDC and VSH analyzes, the fractions of realizations satisfying $h_{\rm rec}>2\sigma$ and $h_{\rm rec}>3\sigma$ remain close to zero throughout most of the investigated amplitude range and increase rapidly only for the largest injected strains $h_c\sim(4-7)\times10^{-11}$, demonstrating a significant loss of sensitivity caused by proper motion selection.

Importantly, this suppression cannot be explained by the removal of GW-induced proper motions themselves, since the expected astrometric GW signal is orders of magnitude smaller than the observed dispersion of quasar proper motions. Instead, the effect arises because the cut $|\mu|<0.1\ {\rm mas\ yr^{-1}}$ preferentially removes quasars with larger uncertainties. The remaining sources are concentrated in regions of the sky with a larger number of \textit{Gaia} observations and therefore become strongly correlated with the \textit{Gaia} scanning law. As a consequence, the sky distribution of quasars becomes significantly more anisotropic than in the full sample, modifying the covariance structure of the proper motion field and violating the assumptions of statistical isotropy underlying both the HDC and VSH analyzes.

These simulations demonstrate that aggressive proper motion cuts introduce a substantial selection bias that systematically lowers the recovered GW amplitudes and reduces the detection efficiency of both methods. The results therefore suggest that the use of strongly truncated quasar samples may lead to overly optimistic upper limits, while simultaneously reducing sensitivity to genuine GW signals. The dominant limitation of current astrometric GW searches is thus not purely statistical noise, but the combined effect of sample selection biases, anisotropic sky coverage, correlated astrometric uncertainties, and residual catalog systematics.
\end{appendix}

\end{document}